\documentclass[aps,prd,showpacs,superscriptaddress,nofootinbib,preprintnumbers,notitlepage,10pt]{revtex4-1} 

\makeatletter
\def\p@subsection{}
\makeatother
\usepackage{aas_macros}
\usepackage{graphicx,amsmath,amssymb,lipsum}

\usepackage{xcolor}
\usepackage{silence}

\definecolor{jade}{rgb}{0.0, 0.66, 0.42}
\definecolor{coralred}{rgb}{1.0, 0.25, 0.25}

\usepackage{enumitem}
\usepackage[colorlinks=true,citecolor=jade,linkcolor=jade,urlcolor=jade, backref=false,pdfborder={0 0 0}]{hyperref}

\usepackage[normalem]{ulem}
\usepackage[T1]{fontenc}
\usepackage[utf8]{inputenc} 
\usepackage{mathtools}
\usepackage{wrapfig}
\usepackage{siunitx}
\usepackage{placeins}
\usepackage{colortbl}
\usepackage{fontawesome5}

\usepackage{caption}
\usepackage{subcaption}
\DeclareCaptionJustification{revjust}{\leftskip=0pt\rightskip=0pt\parfillskip=0pt plus 1fil}
\newcommand{\subfigref}[2]{\hyperref[#2]{#1}}
\newcommand{\invisiblecaption}[1]{
\captionsetup{labelformat=empty}
\caption{}
\label{#1}
\captionsetup{labelformat=simple}
}
\usepackage{calc}

\usepackage{booktabs}

\definecolor{vlightgray}{gray}{0.9}
\usepackage{ulem}
\usepackage{diagbox}
\usepackage{bm}
\usepackage{bbm}

\renewcommand\thesubsection{\arabic{section}.\arabic{subsection}}
\newcommand{\be}{\begin{equation}}
\newcommand{\ee}{\end{equation}}
\newcommand{\beqa}{\begin{eqnarray}}
\newcommand{\eeqa}{\end{eqnarray}}

\newcommand{\bseq}{\begin{subequations}}
\newcommand{\eseq}{\end{subequations}}

\renewcommand{\ln}{\mathop{\rm ln}\nolimits}

\def\gsim{\raise0.3ex\hbox{$\;>$\kern-0.75em\raise-1.1ex\hbox{$\sim\;$}}}
\def\lsim{\raise0.3ex\hbox{$\;<$\kern-0.75em\raise-1.1ex\hbox{$\sim\;$}}}

\def\beqn#1{\begin{equation}\label{#1}}
\def\eeqn{\end{equation}}

\def\beqa#1{\begin{eqnarray}\label{#1}}
\def\eeqa{\end{eqnarray}}

\makeatletter
\newlength{\apb@width}
\newcommand{\autoparbox}[2][c]{\settowidth{\apb@width}{#2}\parbox[#1]{\apb@width}{#2}}

\makeatother

\def\Z2{$\mathcal{Z_2}$}

\newcommand {\ignore}[1]{}

\begin{document}

\preprint{RBI-ThPhys-2025-40}

\title{\texorpdfstring{\large 
Generating optimal Gravitational-Wave template banks with\\ metric-preserving autoencoders}{Autoencoders for GW template banks}}

\author{Giovanni Cabass}
\email{giovanni.cabass@gmail.com}
\affiliation{Division of Theoretical Physics, Ru{\dj}er Bo{\v s}kovi{\'c} Institute, Zagreb HR-10000, Croatia}

\author{Digvijay Wadekar}
\affiliation{\mbox{Department of Physics and Astronomy, Johns Hopkins University,
3400 N. Charles Street, Baltimore, Maryland, 21218, USA}}
\affiliation{\mbox{Weinberg Institute, University of Texas at Austin, Austin, TX 78712, USA}}

\author{Matias Zaldarriaga}
\affiliation{School of Natural Sciences, Institute for Advanced Study, 1 Einstein Drive, Princeton, NJ 08540, USA}

\author{Zihan Zhou}
\affiliation{Department of Physics, Princeton University, Princeton, NJ 08540, USA}

\begin{abstract} 
\noindent Matched filtering for signal detection in noisy data requires template banks that capture variation in signal waveforms while minimizing computational cost. Dimensionality reduction of signal waveforms can be important for building efficient template banks. In various domains of physics, dimensionality reduction is very commonly performed using linear methods such as singular value decomposition (SVD). This can, however, introduce redundancies if the signals span curved, nonlinear manifolds in parameter space. Alternatively, autoencoders are a type of neural networks that can be used for non-linear dimensionality reduction. We use a variation of the autoencoder which preserves the metric in its latent space ($g_{ij}^{\text{latent}} \approx g_{ij}^{\text{physical}}$); this enables template banks to be constructed by simply placing a uniform grid in the autoencoder's low-dimensional latent space. We apply our method for creating geometric template banks for gravitational wave searches and show that our banks require fewer dimensions compared to using the SVD basis. Our method can also be useful for other applications requiring dimensionality reduction, such as gravitational waveform modeling, fast parameter estimation and model-independent tests of general relativity. Finally, we discuss extensions to other domains including cosmological parameter estimation, and we show tests of our method in extreme cases of periodic signal manifolds. \href{https://github.com/giovanni-cabass/autoencoder_GWs}{\faGithub}


\end{abstract}

\maketitle

\section{Introduction} 
\label{sec:intro}

\noindent Template banks are systematically-constructed collections of ``template'' waveforms or patterns used to identify or extract signals embedded in noisy data by matched filtering (comparing the data to each template). Template banks are crucial in signal-detection any field where:
$(i)$ the expected signal shape is parameterizable,
$(ii)$ the data is noisy,
$(iii)$ direct identification is challenging. Template banks have been used in various areas of physics and astronomy, such as Gravitational Wave (GW) data analysis (the main topic of this paper, see e.g.~\cite{Wadekar:2023xbe,Abbott:2016blz,Abbott:2018gqg,LIGOScientific:2020ibl,Sachdev:2019aeb,Usman:2015kfa,Andres:2021jtr,LIGOScientific:2021djp,LIGOScientific:2024nkj,Venumadhav:2019ljl,Olsen:2022gts,Mehta:2023xbe,Chia:2023dsa,Nitz:2019gbi,Nitz:2021bwt,Nitz:2021zqy, LVK_GWTC4, Coo22} as references), Pulsar and Fast Radio Burst (FRB) searches, Exoplanet Transit and Radial Velocity searches, Binary Star/Compact Object searches in X-ray and Gamma-ray Astronomy, Signal Processing in Particle Physics, Cosmic Ray and Air Shower Detection, among many more.

\begin{figure}
\centering
\includegraphics[width = \textwidth]{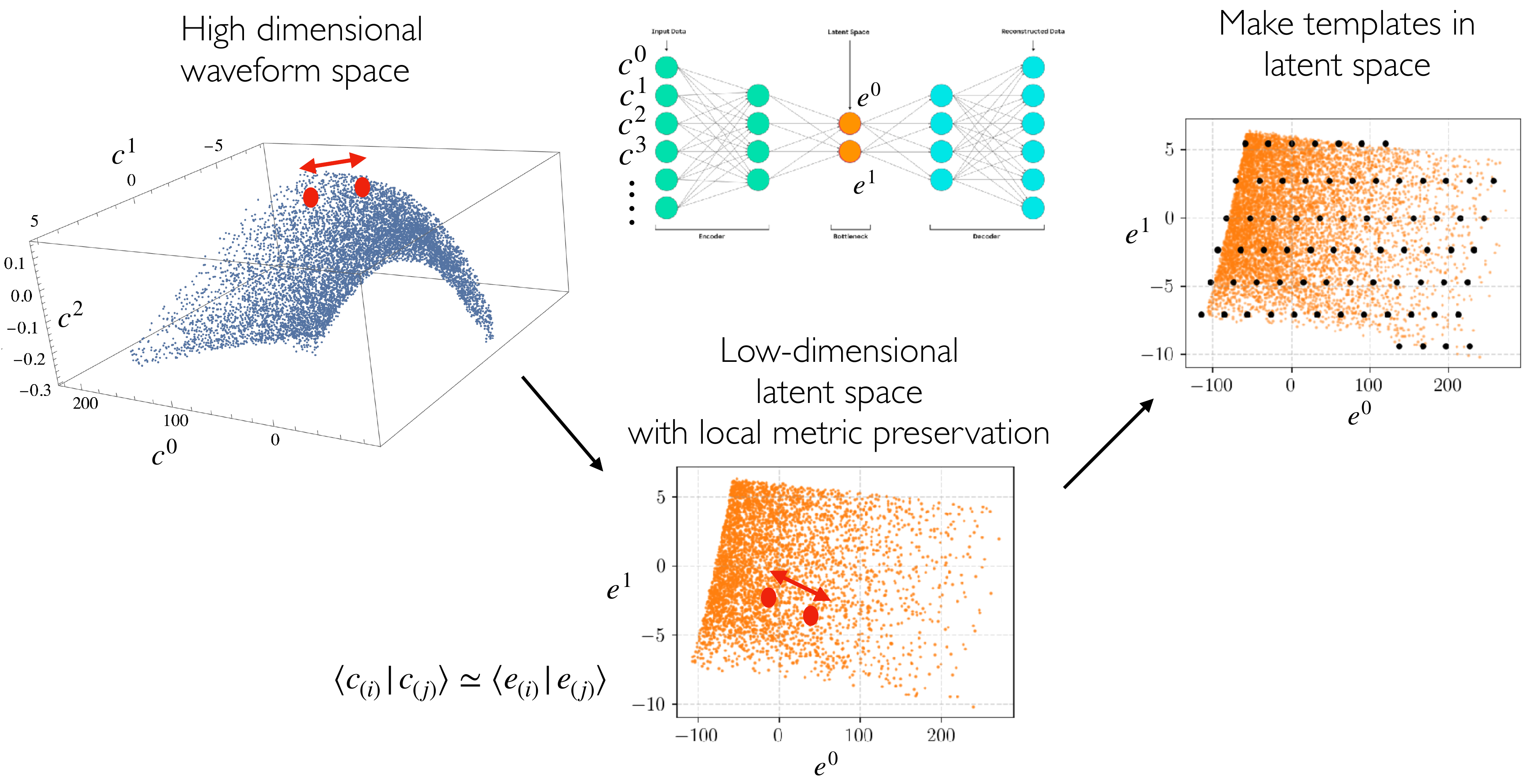} 
\caption{A schematic giving an overview of our procedure for constructing the template banks using metric-preserving autoencoders.}
\label{fig:summary}
\end{figure}

Typically, the signal waveforms are high-dimensional, but most of variation in the waveforms is effectively only along a few dimensions. For example, in GW waveforms, the dominant variation occurs along the dimension of the binary chirp mass, and the next most important variation occurs in a combination of the binary mass ratio and effective aligned spin. The dimensionality of the template bank is set by the number of such effective dimensions.
Thus the problem becomes to capture the variation in signal waveforms into the minimum possible number of dimensions while also retaining accuracy. Apart from being of the right dimensions, the banks should have templates optimally spaced along the dimensions, otherwise this introduces redundancies (as overly-close templates give the same result) and this also increases the search computational cost. This approach goes under the name of geometric placement, see e.g.~\cite{Roulet:2019hzy, Han23_GeometricPlacement, Brown12_GeometricPlacement, Phu24_Ecc_template_banks}.

In various domains of physics and astronomy, dimensionality reduction is commonly performed using linear methods such as singular value decomposition (SVD) or principal component analysis (PCA). However, these methods are linear and therefore cannot capture the non-linear variation in the signal waveforms.
Machine learning architectures like autoencoders are good at non-linear dimensionality reduction; autoencoders are neural network architectures which have a latent bottleneck layer with a smaller number of dimensions than the input (see Fig.~\ref{fig:summary}) \cite{Hinton2006,Kingma2014,Vincent2008, Vincent2010, Bengio2007}. They have been used in a few applications in astronomy and cosmology \cite{Melchior2023,Liang2023a, Liang2023b, Arcelin2021}. The input data is therefore compressed into a latent vector and then reconstructed back to the original dimensions ${\bm c} \rightarrow {\bm e} \rightarrow \hat{{\bm c}}$. The reconstruction loss is based on the difference between ${\bm c}$ and $\hat{{\bm c}}$. The standard autoencoder models do not typically preserve the metric in the input space. We therefore construct metric-preserving autoencoders in this paper using a hybrid loss function given by
$\smash{L = \alpha\times L_{\rm reconstruction} + \beta\times L_{\text{distance preservation}}}$, 
where $\alpha$ and $\beta$ are hyperparameters. We will use the metric preservation loss to enforce $\smash{|{\bm c}-{\bm c}'| \approx |{\bm e}-{\bm e}'|}$.

The main results of the paper are Figs.~\ref{fig:GWs_phase_overlap} and \ref{fig:GWs_effectualness} (that summarize the effectualness of our template banks), together with Tab.~\ref{tab:template_banks} (which shows the number of templates per bank and sub-bank). We also compare our results using autoencoders to those of Ref.~\cite{Wadekar:2023kym} who use a different ML algorithm for dimensionality reduction (random forests). We also compare the performance of the linear and non-linear dimensionality reduction techniques by showing the results of a linear model using only the first few SVD coefficients for the waveforms.

\section{Geometric template banks for Gravitational Wave searches} 
\label{sec:GWs}

Using metric-preserving autoencoders first requires calculation of distance in the physical waveform space using an inner product. Refs.~\cite{Roulet:2019hzy,Wadekar:2023xbe} transformed GW waveforms into a SVD space where the distance between templates is simply the Cartesian distance. We use the PhenomXAS model for simulating the waveforms \cite{Pratten:2020fqn}, which are then used to construct the banks. In this work, we start from the templates banks of \cite{Wadekar:2023xbe}. They divided space of waveforms into different banks by isolating waveforms $h({\bm p},f) = A({\bm p},f)\exp{\rm i}\Psi({\bm p},f)$ that have a similar amplitude profile and placing them in $17$ banks $\text{BBH-0}$ to $\text{BBH-16}$ (${\bm p}$ being the binary's parameters and $f$ the frequency). The banks were further subdivided in sub-banks based on the chirp mass ${\cal M}$, which controls the duration of the signal. Then, one performs a dimensionality reduction for the phases using SVD, in such a way that $\Psi({\bm p},f)$ in each sub-bank can be described by a finite (and small) number of basis functions and respective SVD coefficients:
\be
\label{eq:phase_SVD_decomposition}
\Psi_{22}({\bm p},f)= \mathrm{arg}[h({\bm p},f)] = \langle\Psi_{22}\rangle_\textup{sub-bank}(f) + \sum_{A=0}^{\rm few}c^A({\bm p})\Psi^{\rm SVD}_A(f)\,\,,
\ee
we only model the leading-order quadrupole mode and hence use the $22$ subscript. $\langle\cdot\rangle_\textup{sub-bank}$ denotes that we construct the SVD after subtracting the average over all the examples in a given sub-bank, and ``$\rm few$'' is equal to $9$ in the remainder of this paper. Then, the search proceeds by performing geometric placement of the templates in the much more manageable space of the SVD coefficients $c^A$. More precisely, the construction of the template bank involves creating a uniform grid in the space of the SVD coefficients, inspired by the fact that the mismatch between templates in a given sub-bank is well approximated by the Euclidean distance between the $c^A$:
\be
\langle h(c)|h(c+\delta c)\rangle\approx1-\frac{\delta_{AB}c^Ac^B}{2}\,\,.
\ee
There can however be a redundancy in the SVD coefficients, which can lead to the banks requiring more dimensions than necessary. In this work, we will therefore encode the sub-manifold where the $c^A$ lie ( $\smash{c \overset{\raisebox{-1.5em}{${\scriptstyle \cal E}$}}{\longrightarrow} e \overset{\raisebox{-1.5em}{${\scriptstyle \cal D}$}}{\longrightarrow} c}$ ) and ensure both that close points in the latent space correspond to close points in the embedding space and that a uniform grid of equally-spaced $(e^0,e^1)$ corresponds to equally-distant waveforms in the physical space. In other words, we want to find ``coordinates'' $e^\mu$ in which the overlap metric is flat.

\noindent We first review the case -- already studied in Ref.~\cite{Wadekar:2023kym} -- of using random forests to model the leading SVD coefficients $(c^0,c^1,\dots c^9)$ for the phase profile of the templates in each amplitude bank and chirp mass sub-bank, comparing it to the case where we consider only the first $2$ SVD coefficients $(c^0,c^1)$ and put to $0$ the others. Then we move to using autoencoders to model $(c^0,c^1,\dots c^9)$. 

\begin{figure}[t]
\centering
\includegraphics[width = 0.495\textwidth]{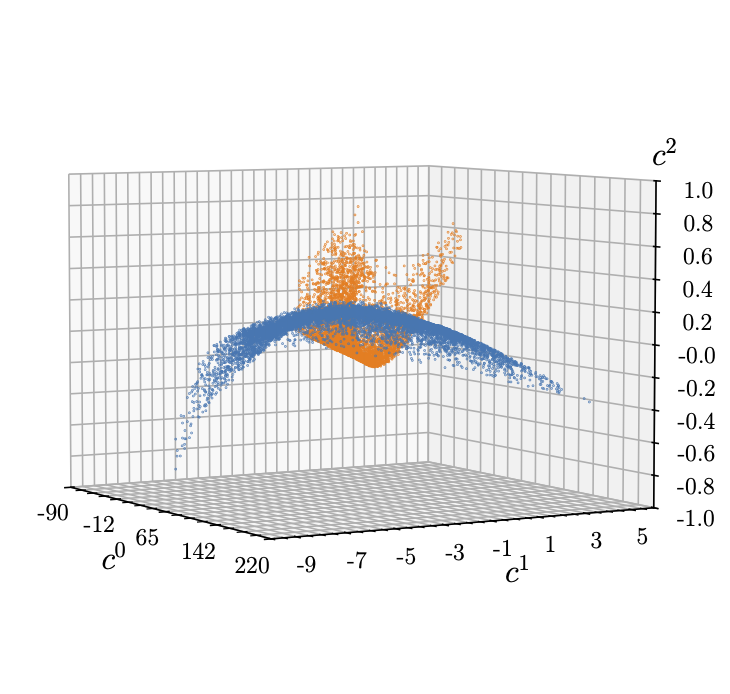} 
\includegraphics[width = 0.495\textwidth]{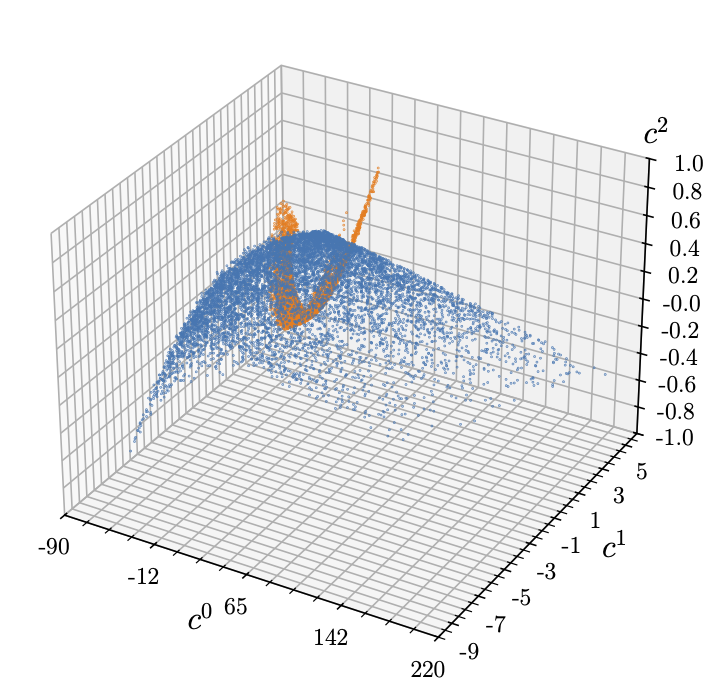} 
\caption{We show the first three SVD coefficients $c^A$ obtained from decomposing the phase of gravitational wave signals: $\Psi_{22}({\bm p},f) = \mathrm{arg}[h({\bm p},f)] =  \langle\Psi_{22}\rangle(f) + \sum_{A=0}^{\rm few}c^A({\bm p})\Psi^{\rm SVD}_A(f)$ for a range of binary parameters ${\bm p}$. We show the coefficients for Bank $0$ (teal) and Bank $4$ (orange). We clearly see that the coefficients follow a narrow curved hypersurface instead of broadly filling the space. Such curved hypersurfaces cannot be compressed efficiently by linear methods like SVD. We will therefore use non-linear dimensionality reduction methods like autoencoders instead. The left and right panels are different views of the same plot.}
\label{fig:3D_stuff}
\end{figure}

Our template banks are listed in Tab.~\ref{tab:template_banks}. Our plots focus on three banks, two at the high-mass and one at the low-mass end: $\text{BBH-0}\, ({\cal M}\in[2.6,5.3]\,{M_\odot}$), $\text{BBH-4}\, ({\cal M}\in[5.8,10.8]\,{M_\odot})$ and $\text{BBH-12}\, ({\cal M}\in[28.4,173.5]\,{M_\odot})$. In the remainder of the text and of the figures, we will denote these by ``Bank $0$'', ``Bank $4$'', ``Bank $12$'' for simplicity. We show the first three SVD components for Bank $0$ and Bank $4$ in Fig.~\ref{fig:3D_stuff}. We see that the manifold of the $c^A$ is not a hyperplane but rather a ``curved'' hypersurface (in other words, it is not a manifold described by a system of inhomogeneous linear equations). Hence we expect that nonlinear dimensionality reduction methods will perform much better than linear ones in reproducing the template waveforms.

\begin{figure}[t]
\centering
\includegraphics[width = 0.99\textwidth]{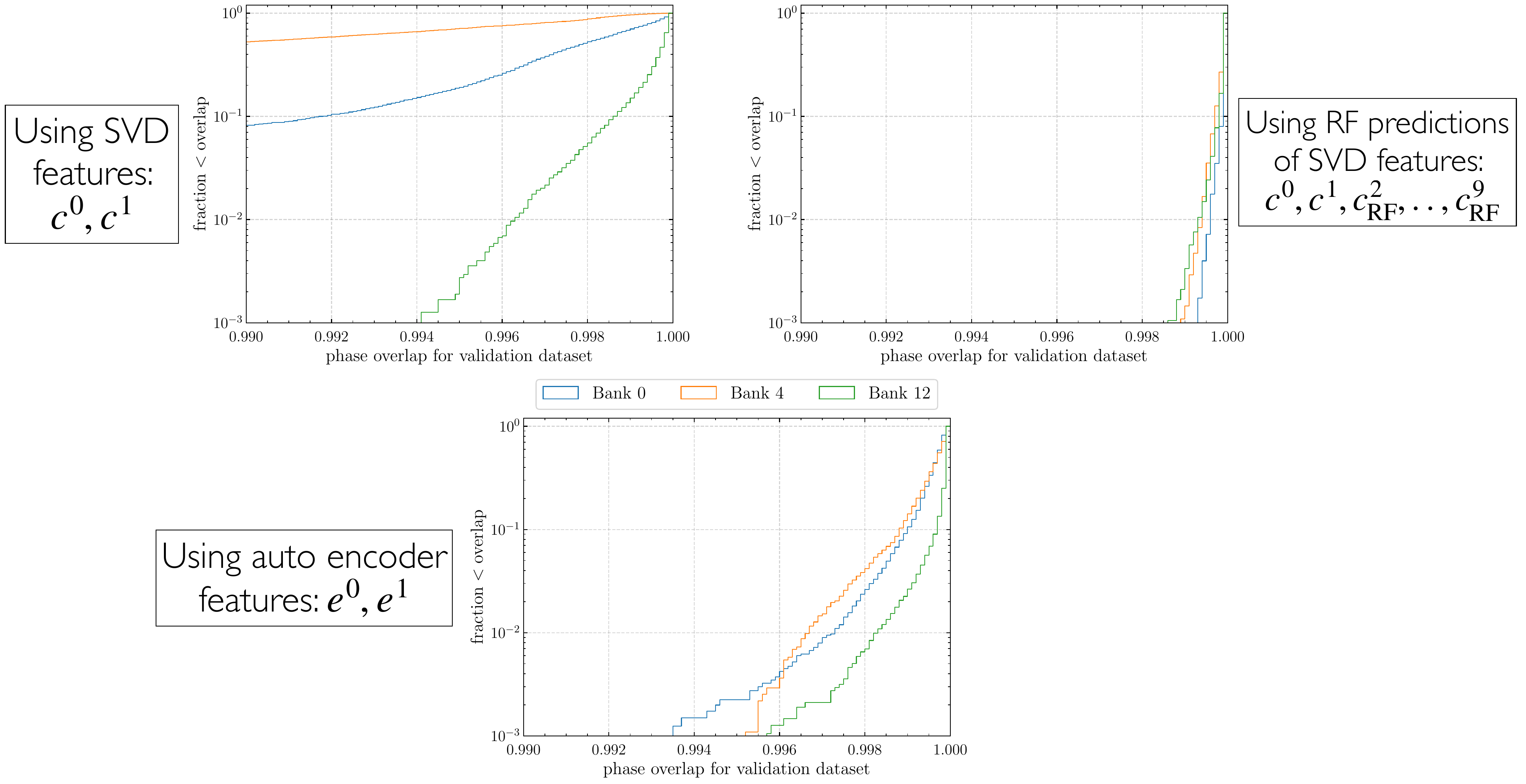} 
\caption{Phase overlap from Eq.~\eqref{eq:phase_overlap} between the validation phases and their different models in the selected banks ${\cal M}\in[2.6,5.3]\,{M_\odot}$ (Bank $0$), $ {\cal M}\in[5.8,10.8]\,{M_\odot}$ (Bank $4$) and $ {\cal M}\in[28.4,173.5]\,{M_\odot}$ (Bank $12$). The upper left panel shows the phase overlap between the validation phases and their model using only the first $2$ SVD basis vectors. In the autoencoder case (lower panel), we see a much lower fraction have reconstruction error larger than $0.1\%$. We also compare our results with another machine learning algorithm, the random forest [RF] (upper right panel), where we predict the supposedly-redundant higher-order SVD coefficients $(c^2,c^3,\dots c^9)$ from the first two $(c^0,c^1)$ \cite{Wadekar:2023kym}. We see that both the machine learning algorithms allow for using an effectively two-dimensional parameter space and significantly outperform using just the linear SVD method, especially for low-mass banks.}
\label{fig:GWs_phase_overlap}
\end{figure}

\begin{figure}[t]
\centering
\includegraphics[width = 0.99\textwidth]{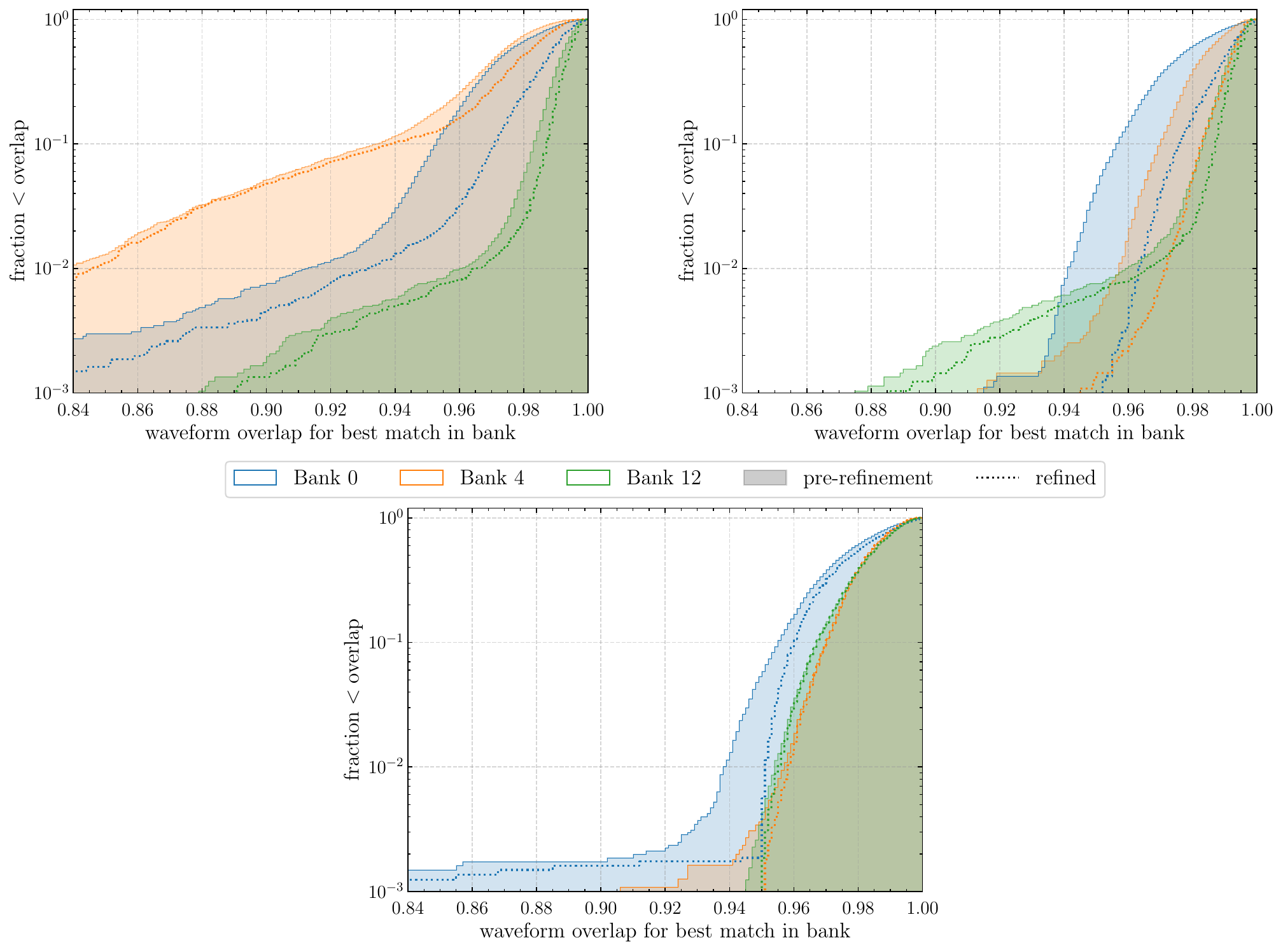} 
\caption{Similar to Fig.~\ref{fig:GWs_phase_overlap}, but showing the effectualness of our banks for the full GW waveform instead of just the phase overlap. We again show the performance of the two dimensional SVD model (upper left panel), autoencoder model (lower panel) and the random forest model (upper right panel). The full bars (dotted lines) show the result before (after) grid refinement (i.e., making the template grid finer). See the main text for details regarding the grid employed for constructing templates in the autoencoder latent space. We again see that both the machine learning algorithms allow for using effectively two-dimensional banks and significantly outperform using just the linear SVD method, especially for low-mass banks.}
\label{fig:GWs_effectualness}
\end{figure}

To compare the performance of our models, we calculate the phase overlap betwen our models and the validation phases using:
\be
\label{eq:phase_overlap}
\text{phase overlap}=\Big|\Big\langle A_{22}^\textup{sub-bank}{\rm e}^{{\rm i}\Psi_{22}}\Big|A_{22}^\textup{sub-bank}{\rm e}^{{\rm i}\Psi_{22}^{\rm model}}\Big\rangle\Big|
\ee
We show the phase overlap between the validation phases and their different models in the selected banks in Fig.~\ref{fig:GWs_phase_overlap}. In the next section, we will attempt to reduce our templates to a two-dimensional model by using autoencoders. 
To benchmark our results, we show in the upper-left panel of that figure the phase overlap between the validation phases and their model using only the first $2$ SVD coefficients $(c^0,c^1)$ (while putting the rest to zero) in Eq.~\eqref{eq:phase_SVD_decomposition}.

\begin{table*}
\centering
\begin{tabular}{p{1.5cm}p{4cm}p{1cm}p{5.5cm}p{5.5cm}}
\hline
Bank & $\mathcal{M}_\textup{bin edges}$ $[M_\odot]$ & $\Delta c$ & N$_\textup{templates}$ in sub-banks (SVD$2$, RF) & N$_\textup{templates}$ in sub-banks (autoencoder) \\ \hline
BBH-$0$ & 2.6, 5.3, 6.8, 8.4, 10.7, 19.8 & 0.55 & 10442, 1702, 774, 465, 280 & 9806, 1621, 724, 422, 250\\
\rowcolor{gray!10}BBH-$1$ & 5.8, 11.9, 15.8, 27.3 & 0.5 & 1902, 318, 157 & 1842, 295, 123\\
BBH-$2$ &8.9, 18.5, 23., 38.6 & 0.45 & 755, 94, 55 & 726, 72, 44\\
\rowcolor{gray!10}BBH-$3$ & 12., 22.4, 27.4, 49.6 & 0.45 & 263, 53, 44 & 240, 36, 33\\
BBH-$4$ &5.8, 10.8, 13.5, 19.5 & 0.4 & 1782, 422, 225 & 1773, 398, 206\\
\rowcolor{gray!10}BBH-$5$ & 14.7, 23.9, 30.3, 64.2 & 0.35 & 174, 67, 63 & 181, 56, 39\\
BBH-$6$ & 10.3, 21., 29.5, 75.9 & 0.3 & 249, 115, 76 & 209, 96, 61\\
\rowcolor{gray!10}BBH-$7$ & 12.9, 30.6, 92.4 & 0.25 & 145, 74 & 114, 74\\
BBH-$8$ & 15.1 39.9, 108.3 & 0.25 & 73, 31 & 47, 31\\
\rowcolor{gray!10}BBH-$9$ & 18.5, 127.3 & 0.35 & 37 & 15\\
BBH-$10$ & 21.1, 149.1 & 0.3 & 19 & 13\\
\rowcolor{gray!10}BBH-$11$ & 24.7, 168.6 & 0.3 & 14 & 7\\
BBH-$12$ & 28.4, 173.5 & 0.3 & 5 & 3\\
\rowcolor{gray!10}BBH-$13$ & 32.6, 173.8 & 0.3 & 6 & 2\\
BBH-$14$ & 37.8, 173.6 & 0.3 & 3 & 1\\
\rowcolor{gray!10}BBH-$15$ & 43.2, 173.7 & 0.25 & 3 & 1\\
BBH-$16$ & 51.8, 166. & 0.2 & 3 & 1\\ \hline
Total & & & 20890 & 19562\\
\hline
\end{tabular}
\caption{Details of the quadrupole-mode gravitational wave template banks. The banks are split according to the normalized amplitude profile -- see Fig.~4 of \cite{Wadekar:2023kym} -- and the sub-banks are split according to the chirp mass ($\mathcal{M}$). The lower and upper bin edges in $\mathcal{M}$ are given in the second column. From BBH-9 onwards, we only have a single sub-bank per bank. $\Delta c$ denotes the grid size in the $(c^0,c^1)$ space that is used to generate the templates. The last two columns show the number of templates we have according to the technique we use for the dimensionality reduction of the space of SVD coefficients. The second-to-last column shows the number of templates we have if we switch on only the $2$ lowest-order SVD coefficients (``SVD$2$'') and use random forests to model the higher-order ones. The last column instead shows the number of templates we have if we use autoencoders with the method discussed in Section~\ref{subsec:autoencoders_GWs}. The number of templates in the two cases is comparable, with the autoencoder resulting in a slightly lower number.}
\label{tab:template_banks}
\end{table*}

\section{Dimensionality reduction with autoencoders}
\label{subsec:autoencoders_GWs}

\begin{figure}[t]
\centering
\includegraphics[width = 0.7\textwidth]{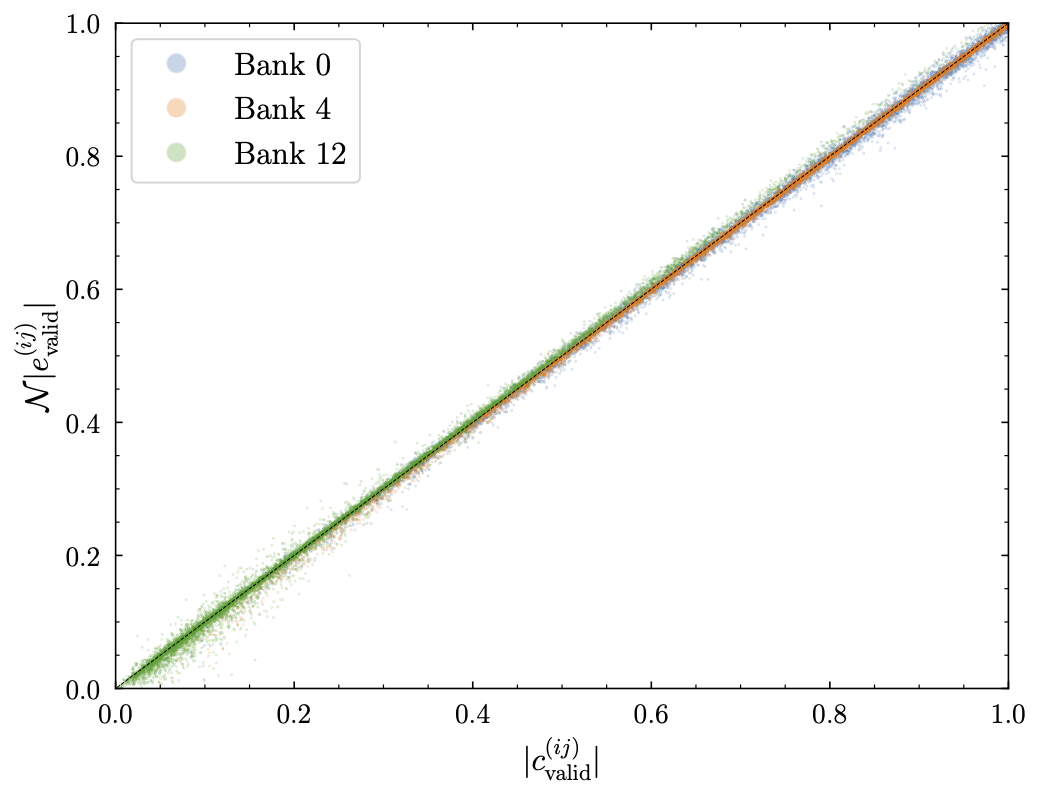} 
\caption{Scatter plot of the distances in the autoencoder latent space (y-axis) vs.~the Euclidean distances in the SVD space (x-axis) for $1\%$ of the pairs of validation points for the three banks discussed in Section~\ref{subsec:autoencoders_GWs}. We see that the distances are preserved extremely well in this case.} 
\label{fig:GWs_autoencoder_distance_preservation_b_0_4_12}
\end{figure}

\noindent We train autoencoders to model the phases of waveforms in each bank and sub-bank. It is possible to use the full high-dimensional GW waveform for training the autoencoder, but it can make the training expensive. We therefore train the autoencoder on a pre-compressed version of the waveform (i.e., we use the top 10 $c^A$ coefficients to represent the waveform). Our goal in this paper is to probe if further compression is possible compared to just using SVD. We follow these steps for the training the autoencoder:
\begin{itemize}[leftmargin=*]
    \item we rescale the data $c^A$, $A=0,\dots 9$ by the standard deviation of the first SVD component $c^0$ (we rescale both training and validation data, and we use half the phases in each sub-bank for training and half for validation). It is generally advisable to rescale the input data to zero mean and unit standard deviation as it helps with efficient training of neural networks: this is our motivation behind rescaling $c^A$. We call this normalization factor $\cal N$.
    \item we construct in \href{https://pytorch.org/}{\texttt{pytorch}} an autoencoder with the architecture $\smash{10 \overset{\raisebox{-1.5em}{${\scriptstyle \cal E}$}}{\longrightarrow} 2 \overset{\raisebox{-1.5em}{${\scriptstyle \cal D}$}}{\longrightarrow} 10}$ to learn the manifold of the $\{d^A\} = \{c^A/{\cal N}\}$. We consider a simple architecture where $\cal E$ and $\cal D$ are both MLPs (Multi-Layer Perceptrons) with $4$ hidden layers, each with $64$ neurons activated by the Gaussian Error Linear Unit (GELU) function. Our model for the manifold of the proper SVD components $\{c^A\}$ is then given by
    \be
    \label{eq:c_A_d_A_definition}
    c^A_{\rm model} = {\cal N}\,{\cal D}^A(e^0,e^1)\,\,,
    \ee
    \item we train the autoencoder by imposing the preservation of distances between embedding and latent space, i.e.~we have $\smash{L = \alpha\times L_{\rm reconstruction} + \beta\times L_{\text{distance preservation}}}$. We use the following loss functions (we denote $\smash{{\cal D}^A({\cal E}({d}_{\rm train}))}$ by $\smash{\tilde{d}^A_{\rm train}}$ here):
    \begin{align}
        L_{\rm reconstruction} &= {\rm MSE}(d_{\rm train},\tilde{d}_{\rm train}) \\
        L_\textup{distance preservation} &= \frac{2}{N_{\rm train}(N_{\rm train}-1)}\sum_{i<j} {\rm e}^{-\frac{|{d}^{(ij)}_{\rm train}|}{\gamma}}\big(|{d}^{(ij)}_{\rm train}| - |e^{(ij)}_{\rm train}|\big)^2\,\,, \label{eq:distance_preservation_loss_GWs}
    \end{align}
    where MSE is the usual mean squared error reconstruction loss and 
    \be
    |{d}^{(ij)}_{\rm train}| = |{d}^{(i)}_{\rm train}-{d}^{(j)}_{\rm train}|\qquad\text{and}\qquad|e^{(ij)}_{\rm train}| = \sqrt{\delta_{\mu\nu}\big[{\cal E}^\mu({d}^{(i)}_{\rm train})-{\cal E}^\mu({d}^{(j)}_{\rm train})\big]\big[{\cal E}^\nu({d}^{(i)}_{\rm train})-{\cal E}^\nu({d}^{(j)}_{\rm train})\big]}\,\,.
    \ee
    We train the autoencoders for all banks and sub-banks using $\alpha=\beta=1$ in the above loss equation. Since we are interested in preserving only local distances, we choose $\gamma=1$ (recall that the goal is to find a coordinate system where the metric is as close as possible to being flat across the whole manifold: in order to find this coordinate system -- assuming that it exists -- it is sufficient to impose that only the distance between neighboring points is preserved).\footnote{When investigating different choices of hyperparameters for Bank $0$, we tried optimizing the choice of $\gamma$ using Simulated Annealing, with an exponential cooling schedule for the temperature ($\smash{T_k=T_0\cdot\texttt{cooling\_rate}^k}$, with starting temperature of order of the validation loss of an initial pre-annealing run with $\gamma=1$), and a log-uniform perturbation to propose candidate $\gamma$ values ($\smash{1/\gamma_{\rm new} = 1/\gamma_{\rm current}+10^U}$, where $U\sim \text{Uniform}({{-2},1})$). Even experimenting with different cooling rates and number of epochs for the annealing, we have not found significant improvements in the distance preservation. Hence, we fixed $\gamma=1$ throughout all our trainings for all banks and sub-banks.} 
    
    We compare the preservation of distances in the embedding and latent spaces in Fig.~\ref{fig:GWs_autoencoder_distance_preservation_b_0_4_12}. We show ${\cal N}|e^{(ij)}_{\rm valid}|$ on the $y$ axis and, on the $x$ axis, we plot $\smash{|{c}^{(ij)}_{\rm valid}| = |{c}^{(i)}_{\rm valid}-{c}^{(j)}_{\rm valid}|}$. We find that distances are preserved extremely well. For each of the three banks we plot only $10000$ randomly-chosen pairs of points among those having a distance $\smash{|{c}^{(ij)}_{\rm valid}|\leq 1}$ in the embedding space. This is done only to allow the reader to distinguish the three colors in the plot: including all pairs would still show the points clustering extremely close to the diagonal. A straightforward check (whose result we do not show here because it would be redundant with Fig.~\ref{fig:GWs_autoencoder_distance_preservation_b_0_4_12},) is to check that the induced metric 
    \be
    \label{eq:induced_metric_GWs}
    g_{\mu\nu}={\cal N}^2\delta_{AB}\frac{\partial{\cal D}^A}{\partial e^\mu}\frac{\partial{\cal D}^B}{\partial e^\nu}
    \ee
    is extremely well approximated as $g_{\mu\nu}\approx{\cal N}^2\delta_{\mu\nu}$.
\end{itemize}

We construct the actual templates in the autoencoder case by adapting the geometric placement method of \cite{Roulet:2019hzy}. We first construct a uniform grid in the embedding space: since we know that the induced metric is very well approximated by ${\cal N}^2\delta_{\mu\nu}$. In order to achieve a spacing $\Delta c$ in the embedding space, we need our grid to have spacing $\Delta e = \Delta c/{\cal N}$. The grid is centered at $\smash{{\cal E}(d^0=0,d^1=0,\dots d^9=0)}$, which corresponds to the mean phase across the sub-bank. We perform the effectualness test for the bank using test waveforms which are different from those used for training the model. For each test waveform, we first extract its SVD components and encode them. We then find the closest grid point in the latent space; the best-fit template to the test signal is the one obtained by decoding this grid point. We trained the autoencoders for all banks and sub-banks in \texttt{torch.float32} precision (which can be refined if needed in the future).


We show the reconstruction error of the autoencoder model from Eq.~\eqref{eq:phase_overlap} in the lower panel of Fig.~\ref{fig:GWs_phase_overlap} and find that it outperforms the 2D SVD case discussed in the last section. The only remaining problem to address is how to determine which grid points in the autoencoder latent space represent physical waveforms and remove the unphysical ones. In order to do this, we train a simple normalizing flow in \href{https://zuko.readthedocs.io/stable/}{\texttt{zuko}}\footnote{We use an unconditional Neural Spline Flow (NSF) \cite{durkan2019neuralsplineflows} with $2$ input features and $4$ layers of transforms, each transform's internal neural network having $2$ hidden layers of $64$ neurons. Notice that the NSF spline transformations are defined over the domain $[-5,5]$ (see \href{https://zuko.readthedocs.io/stable/api/zuko.flows.spline.html}{here} for the \texttt{zuko} implementation): since our features lie in this box for all banks and sub-banks we do not standardize features before training.} for each bank to predict the distribution of points in the latent space (the encoding of the examples we used to construct our $c^A$ dataset). We then drop all grid points that are outside the $99.7\%$ CL contour of this distribution. The last column of Tab.~\ref{tab:template_banks} lists the number of templates obtained with this procedure, while Fig.~\ref{fig:bank_points} shows an example of the latent- (top panel) and SVD- (bottom panel) space distributions of points for Bank $0$ and Bank $4$. 
We show the effectualness test of our banks in the lower panel of Fig.~\ref{fig:GWs_effectualness}.
We also show the results of a refinement procedure by constructing a new template grid in the latent space with a smaller grid spacing. 

\begin{figure}[ht]
    \centering
    \newlength{\FigHeight}
    \settoheight{\FigHeight}{\includegraphics[width=0.99\textwidth]{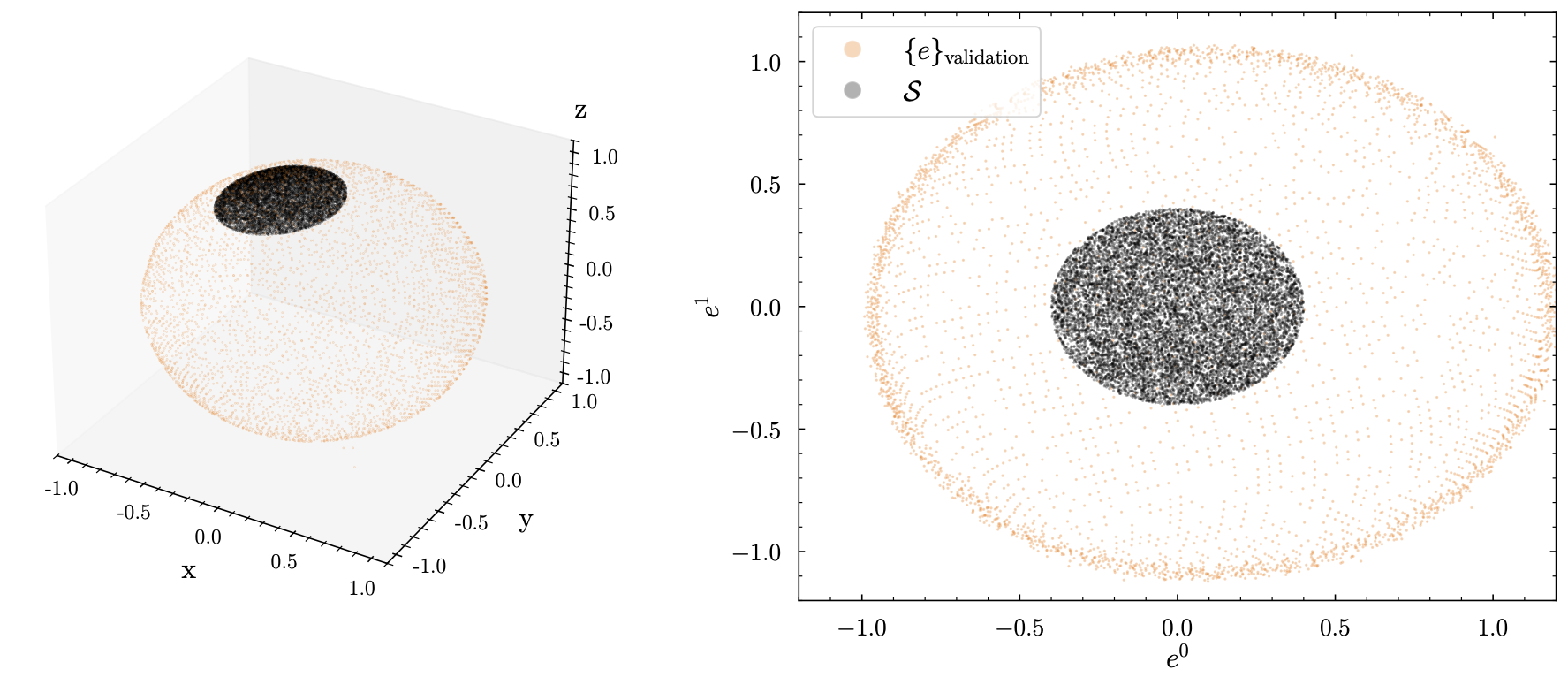}}
    \begin{subfigure}{\textwidth}
        \centering
        \resizebox{!}{\FigHeight}{\includegraphics[width=\textwidth]{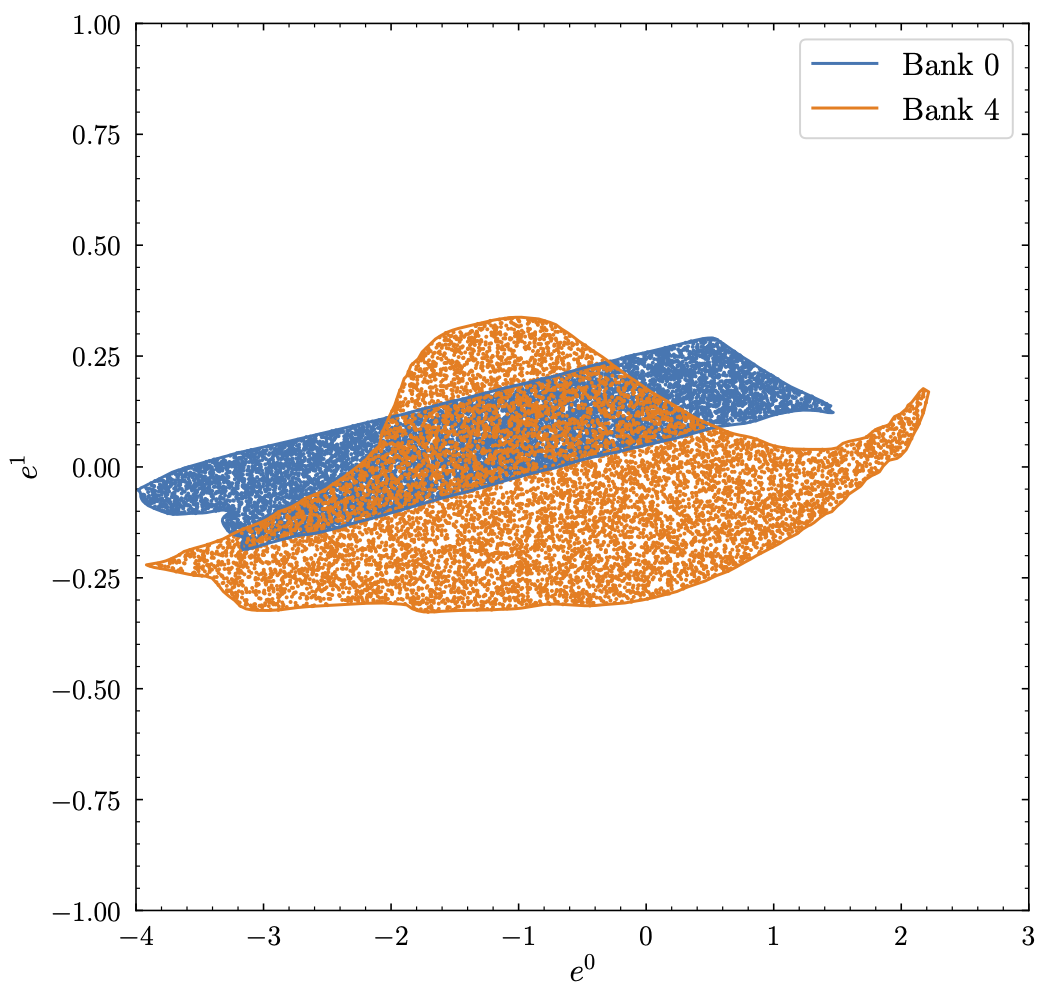}}
        \caption{\small To visualize the latent space of the autoencoders, we show a collection of $10000$ random points within the $99.7\%$ CL contour of the distribution of latent-space points for Bank $0$ and Bank $4$.}
        \label{fig:bank_points_embedding_space}
    \end{subfigure}
    \hfill
    \begin{subfigure}{\textwidth}
        \centering
        \includegraphics[width=0.495\textwidth]{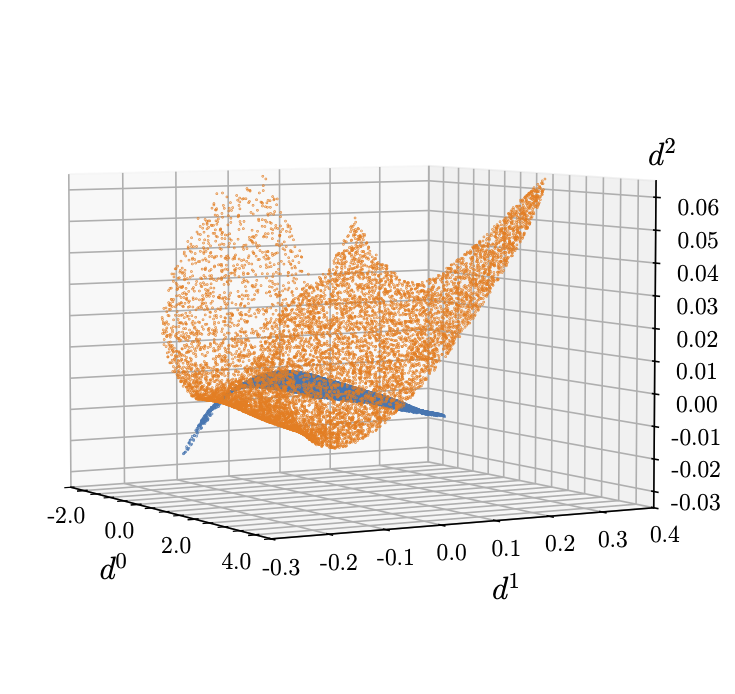}
        \includegraphics[width=0.495\textwidth]{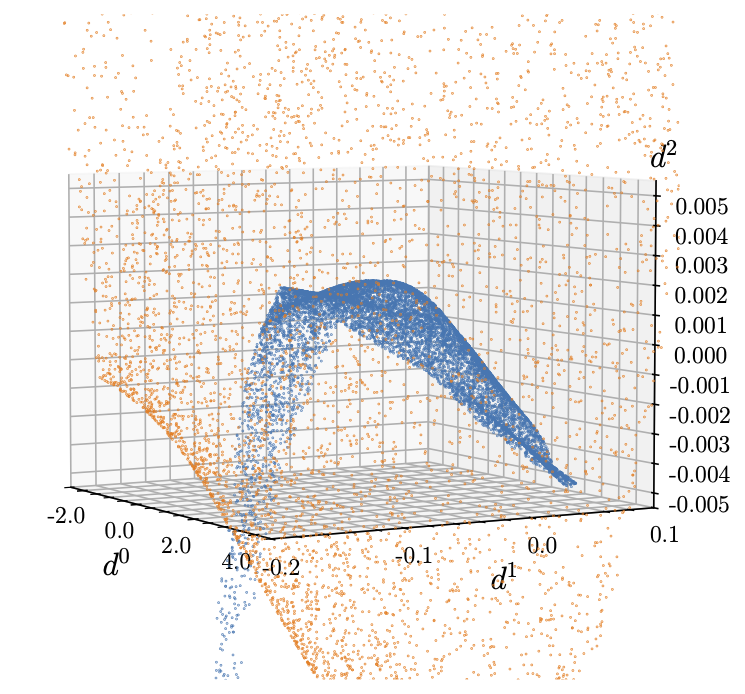}
        \caption{\small We project the latent-space points to the original SVD space using the decoder network to visualize the reconstructed distribution. The first three components of the decoded $\smash{d^A}$ (see Eq.~\eqref{eq:c_A_d_A_definition} for their definition) for the $10000$ latent-space points shown in Fig.~\subfigref{\ref{fig:bank_points}a}{fig:bank_points_latent_space}. The right panel is a zoomed-in version of the left panel.}
    \label{fig:bank_points_latent_space}
    \end{subfigure}
\invisiblecaption{fig:bank_points}
\vspace*{-\baselineskip}
\end{figure}

Before concluding, it is interesting to discuss the curvature of the banks. Post-facto, we can conclude that Bank $0$, Bank $4$ and Bank $12$ have all negligible curvature, since for all of them we were able to find a distance-preserving coordinate chart and the induced metric in each case is very well approximated by ${\cal N}^2\delta_{\mu\nu}$. Because our isometric embeddings have high codimension, our method does not impose specific limits on the intrinsic curvature of the manifolds we can model. However, we find that all our banks and sub-banks have small or negligible curvature. We discuss this further in Appendix~\ref{app:autoencoder_curvature}.

\section{Dimensionality reduction with random forests}

Let us now compare our results to the methodology of Ref.~\cite{Wadekar:2023kym} who used a different ML algorithm for dimensionality reduction, i.e., random forest regressor (RF). We use the RF method implemented in the IAS pipeline code\footnote{The IAS pipeline code is publicly available at \url{https://github.com/JayWadekar/gwIAS-HM}.}.
We first construct the same SVD decomposition described earlier in section~\ref{sec:GWs}. We use the $\rm RF$ (we use \url{https://scikit-learn.org/stable/modules/generated/sklearn.ensemble.RandomForestRegressor.html} with $50$ trees and maximum depth of $50$) to predict the function: $(c^2,c^3,\dots c^9)=f(c^0,c^1)$. In the upper-right panel of Fig.~\ref{fig:GWs_phase_overlap}, we show the phase reconstruction error using Eq.~\eqref{eq:phase_overlap} in the validation dataset for this RF.

For testing the effectualness of the RF method, we again construct the templates using a uniform grid in the $(c^0,c^1)$ space with the grid spacing $\Delta c$ is chosen in the same way as in \cite{Roulet:2019hzy,Wadekar:2023kym}, and listed in the third column of Tab.~\ref{tab:template_banks}. In Fig.~\ref{fig:GWs_effectualness}, we show the effectualness tests for the same three banks of Fig.~\ref{fig:GWs_phase_overlap}. For each test waveform, we first unwrap its phase and take the dot product with the phase basis functions of Eq.~\eqref{eq:phase_SVD_decomposition} to find the $(c^0_{\rm test},c^1_{\rm test})$ coefficients of this test waveform. We then find the closest grid point $(c^0_{\rm closest},c^1_{\rm closest})$ in the $(c^0,c^1)$ space. We use this grid point to construct the best-fit template in the bank using Eq.~\eqref{eq:phase_SVD_decomposition}, in particular, we use $(c^0,c^1)=(c^0_{\rm closest},c^1_{\rm closest})$ and $(c^2,c^3,\dots c^9)={\rm RF}(c^0_{\rm closest},c^1_{\rm closest})$. We calculate the match of the best-fit template with the test waveform and report the result as effectualness. For reference, the top left panel of Fig.~\ref{fig:GWs_phase_overlap} shows the comparison of the case without the RF (where $(c^2,c^3,\dots c^9)$ are simply put to zero), and we indeed find a large improvement due to the RF.

Similar to the previous section, we refine our effectualness calculation by matching the test waveforms to their closest grid points on a new and refined grid with spacing $\Delta c\to\Delta c/2$. Note however that, along the $(c^0,c^1)$ dimensions, not all the points of the rectangular grid describe physically viable waveforms. We use the method adopted in \cite{Roulet:2019hzy} to reject the grid points which do not have a physical waveform in their vicinity (see Fig.~4 of \cite{Roulet:2019hzy} for reference). The second column of Tab.~\ref{tab:template_banks} shows the number $\smash{N_{\rm templates}}$ of templates, i.e.~how many of these grid points are ``physical'' for each bank and sub-bank. We find that our results are comparable to those from the autoencoder and both the machine learning algorithms outperform the linear SVD model. In Appendix~\ref{app:sphere} and Fig.~\ref{fig:sphere_autoencoder_vs_RF}, we show toy examples of cases where using autoencoders can significantly perform the RF results, however for the case of realistic GW examples, we find that both RF and autoencoder methods have comparable accuracy throughout the parameter space that we tested.


\section{Discussion}
\label{sec:Discussion}
Our geometric placement method with autoencoders relies on us being able to find a coordinate system where the metric is equal to ${\cal N}^2\delta_{\mu\nu}$ across the whole manifold. This might not be physically possible if the waveform space has non-zero curvature, comparable to the grid spacing $\Delta c/{\cal N}$ (unlike the gravitational wave cases we encountered where the waveform space is nearly flat). In this case an alternative, albeit much more complex to implement, direction one could investigate is the following:
\begin{enumerate}[leftmargin=*]
    \item we first train the autoencoder to reach a given level of precision in the reconstruction of the manifold;
    \item since we have the possibility of computing the distance $D(e_{(i)},e_{(j)})$ between any two points on the manifold by solving the geodesic equation, we can construct for a set of random points $\{e_{(i)}\}$ the functional 
    \be
    \label{eq:functional}
    \sum_{i<j}\frac{1}{D(e_{(i)},e_{(j)})^\alpha}\,\,;
    \ee
    \item minimizing this functional should organize the points $\{e_{(i)}\}$ in a ``grid'' that is likely not uniform, but is such that all neighboring points have around the same geodesic distance from one another.
\end{enumerate}
It could be beneficial, after having trained the autoencoder, to find a coordinate change $e=e(e')$ (i.e.~$\smash{c_{\rm model}\propto{\cal D}(e(e'))}$ is the decoder now) such that the $e'$ lie on a domain that is easier to deal with than those seen e.g.~in Fig.~\ref{fig:autoencoder_curvature}. This can be achieved with the invertible transformations typically used to train normalizing flows. Even then, successfully implementing this method requires us to deal with a lot of unknown variables. For example, the choice of what power $\alpha$ of the inverse geodesic distance to put in Eq.~\eqref{eq:functional}, and the fact that it is not known in advance at which distance the points will stabilize (this likely being a function of how many points we start with). Furthermore, the fact that computing the geodesic distance on curved manifolds is not an easy task in general and an area of active research \footnote{Notice that, given that our metric is already in \texttt{pytorch}, one could try to compute $\smash{D(e_{(i)},e_{(j)})}$ by minimizing over $\Phi$ the functional
\begin{equation}
\frac{1}{2}\int_0^1{\rm d}\lambda\,g_{\mu\nu}\big(e(\lambda)_\Phi\big)\frac{{\rm d}e^\mu(\lambda)_\Phi}{{\rm d}\lambda} \frac{{\rm d}e^\nu(\lambda)_\Phi}{{\rm d}\lambda}\,\,,
\end{equation}
where $\smash{e^\mu(\lambda)_\Phi=e_{(j)}^\mu\lambda + (1-\lambda)e_{(i)}^\mu + \lambda(1-\lambda){\rm MLP}^\mu(\lambda)_\Phi}$. Finding the optimal weights $\Phi$ of the MLP, initializing them such that the ${\rm MLP}$ vanishes, will find both the geodesic curve and, from it, $\smash{D(e_{(i)},e_{(j)})}$. Knowledge of the Christoffel symbols and the formulas of Appendix~\ref{app:sphere} could be used to inform better the form of $\smash{e^\mu(\lambda)_\Phi}$ at the boundaries and accelerate convergence. Moreover, we notice that having the Christoffel symbols in \texttt{pytorch} as in Appendix~\ref{app:sphere} allows us to perform random geodesic walks on the manifold with a constant -- albeit small -- step, such that every point is at a fixed geodesic distance from the preceding one \cite{Schwarz_2023}. However these points will not be distributed uniformly but according to the volume form on the manifold and, most importantly, this would not be a geometric but a stochastic placement of templates.}, we leave exploring this alternative direction to future work.

\begin{figure}[t]
\centering
\includegraphics[width = 0.495\textwidth]{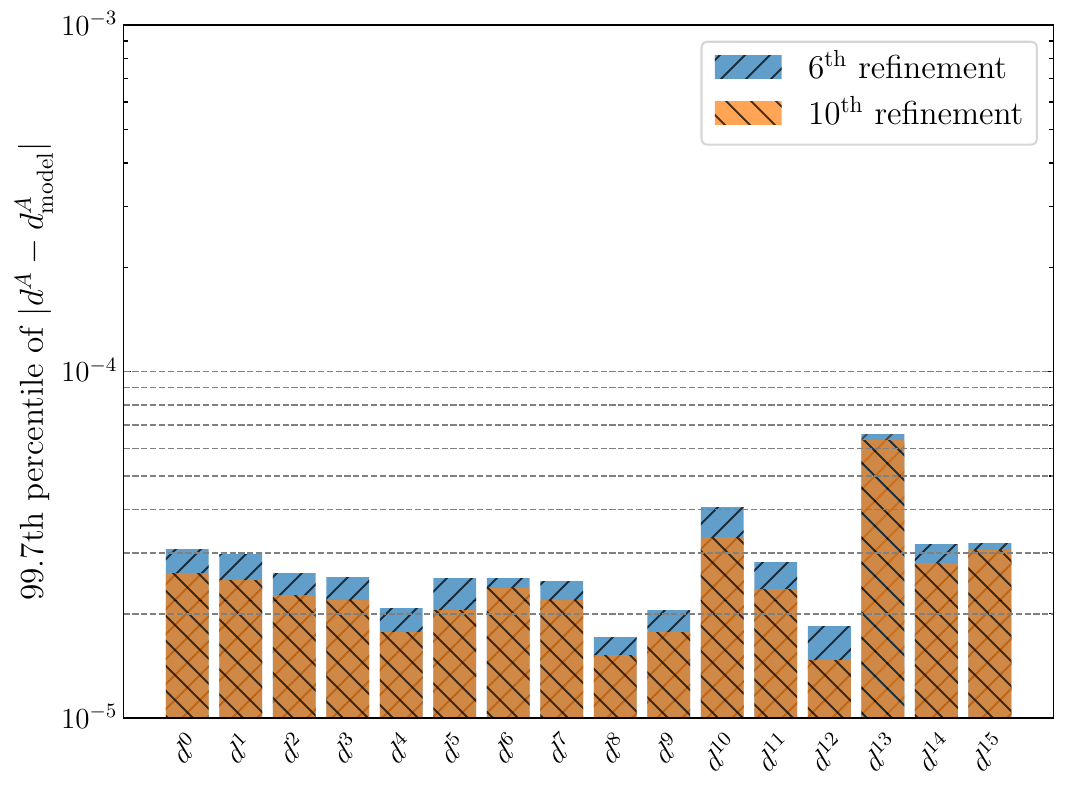} 
\includegraphics[width = 0.495\textwidth]{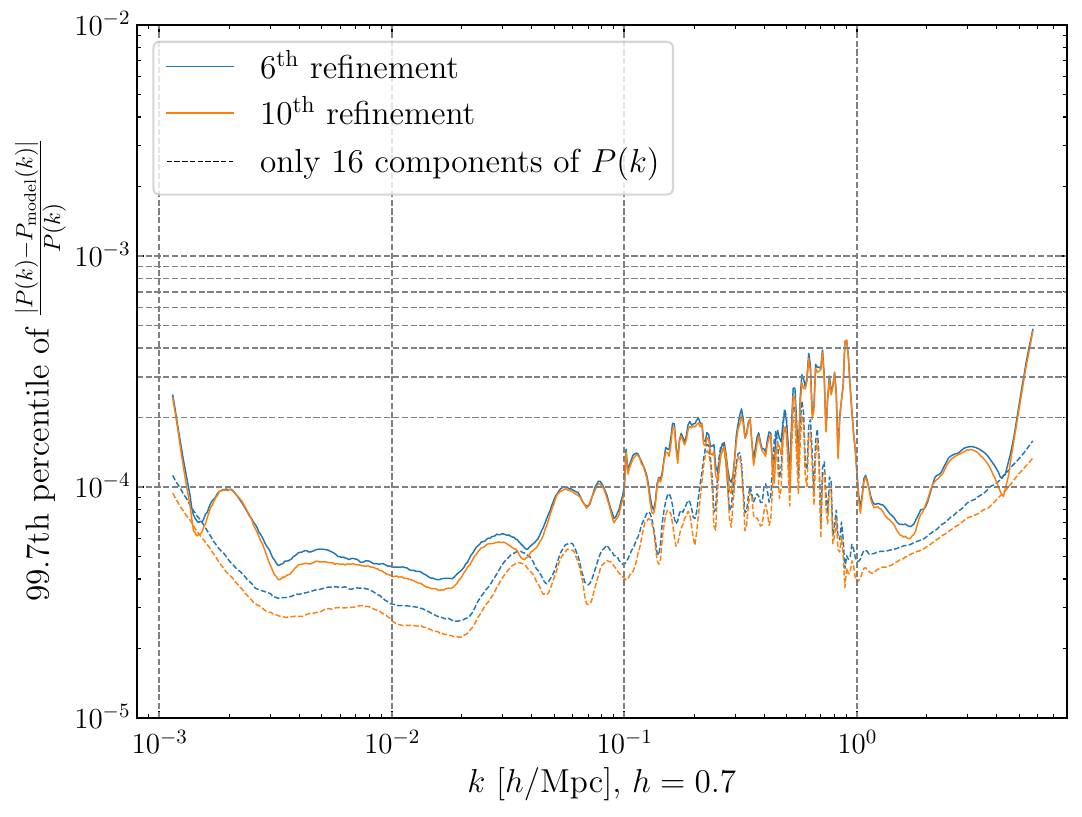} 
\caption{We use our method to perform dimensionality reduction of the linear matter power spectrum in cosmology. Our decoder $\smash{\cal D}$ aims at reconstructing the rescaled SVD coefficients $\smash{d^A=c^A/{\cal N}}$. We use the knowledge distillation technique \cite{hinton2015distillingknowledgeneuralnetwork} to train our network. We first trained $5$ teachers using the simple reconstruction loss $\smash{{\rm MLP}(d_{\rm train},\tilde{d}_{\rm train})}$, where $\smash{\tilde{d}^A_{\rm train} = {\cal D}(\theta_{\rm train})}$ and $\theta_{\rm train}$ are the cosmological parameters $(\omega_{\rm c},\omega_{\rm b},n_{\rm s})$ (each batch of the training dataset contains a random selection of cosmological parameters from the regular grid used to generate the training spectra). Then, we trained a student with the same architecture as the $5$ teachers using transfer learning: the loss was the sum of a reconstruction loss $\smash{{\rm MLP}(d_{\rm train},\tilde{d}^{\rm student}_{\rm train})}$ and $\smash{\smash{{\rm MLP}(\langle d^{\rm teacher}_{\rm train}\rangle,\tilde{d}^{\rm student}_{\rm train})}}$, where $\smash{\langle d^{\rm teacher}_{\rm train}\rangle}$ is the mean of the predictions of the $5$ teachers. Following the initial transfer learning procedure, the student model was iteratively fine-tuned for ten subsequent iterations. The left panel shows the $99.7$th percentile of the absolute error for each of the $16$ $d^A$ in the validation set, for the fifth-to-last and last refinement. The right panel shows instead the $99.7$th percentile of the relative error in the validation power spectra as a function of wavenumber. By comparing the full with the dashed lines (the dashed lines showing the error with respect to the validation spectra where all the SVD components after the $16^{\rm th}$ are set to zero) shows that we are mostly dominated by the error coming from us keeping only the first $16$ basis functions.} 
\label{fig:cobra}
\end{figure}

\subsection{Application to modeling the cosmological matter power spectrum}
\label{sec:Discussion:cosmology}

Let us discuss the possibility of performing dimensionality reduction using autoencoders in the space of another set of SVD coefficients, this time in the context of cosmology. More precisely, we have in mind the \texttt{COBRA} decomposition of the linear matter power spectrum of Ref.~\cite{Bakx:2024zgu}. One can generate a set of linear power spectra as a function of a number of cosmological parameters ($\omega_{\rm c}$, $\omega_{\rm b}$ and $n_{\rm s}$, for example), and construct its SVD to find a small number of basis functions with which to represent any linear power spectrum. More precisely, following Ref.~\cite{Bakx:2024zgu} we generate power spectra at $500$ wavenumbers between $\smash{0.8\times10^{-4}\,{\rm Mpc}^{-1}}$ and $\smash{4\,{\rm Mpc}^{-1}}$ on a regular ($40\times20\times20$) grid $(\omega_{\rm c},\omega_{\rm b},n_{\rm s})\in[0.08,0.175]\times[0.020,0.025]\times[0.8,1.2]$ using \href{https://camb.readthedocs.io/en/latest/}{\texttt{CAMB}} (see the last column of Tab.~1 of \cite{Bakx:2024zgu} for more details), and we extract the first $16$ coefficients ($c^A$) as in Ref.~\cite{Bakx:2024zgu} (with the difference that besides rescaling the power spectra by the mean of the templates, we also subtract the mean of these rescaled spectra before performing the SVD): this is our training set.

We then uniformly generate $32256$ random validation spectra in the same $\omega_{\rm c}$, $\omega_{\rm b}$ and $n_{\rm s}$, range. There are two differences with the GW case discussed in this paper. First, now we would like that the latent space is the space of cosmological parameters $(\omega_{\rm c},\omega_{\rm b},n_{\rm s})$, effectively finding the ``inverse'' of the Boltzmann code used to generate the templates. This is less trivial than just autoencoding the $c^A$ as done in this paper. Second, Ref.~\cite{Bakx:2024zgu} only generated templates on an uniform and structured grid, without a notion of distance between the resulting spectra or the space of the $c^A$, unlike e.g.~what was discussed in Ref.~\cite{Philcox:2020zyp} in the context of cosmology. Hence, we are not interested -- when comparing with Ref.~\cite{Bakx:2024zgu} -- in constructing metric-preserving autoencoders. In Fig.~\ref{fig:cobra} we show a preliminary result for what we could call the training of the decoder only: we construct a simple MLP with a $3$-dimensional input layer, $8$ hidden layers each containing $128$ neurons and a GELU activation, and a $16$-dimensional output layer, and train it to reconstruct $d^A=c^A/{\cal N}$ where $\cal N$ is defined as in Section~\ref{subsec:autoencoders_GWs}.\footnote{Unlike Section~\ref{subsec:autoencoders_GWs}, here $\cal N$ is defined using only training data.} We refer to the caption of Fig.~\ref{fig:cobra} for more details about the training. While our result can already be used as an alternative to the Radial Basis Function (RBF) interpolation of \cite{Bakx:2024zgu} -- we checked that it has comparable speed (e.g.~we can predict our $32256$ validation spectra in $\sim 660\,{\rm ms}$ on a single core) while having slightly better precision (compare the magenta line in the lower panel of Fig.~1 of \cite{Bakx:2024zgu} with the right panel of Fig.~\ref{fig:cobra}) -- it will be interesting to check if we can construct an autoencoder for the $16$ SVD coefficients using this MLP as starting point for the decoder, and to investigate in which sense a distance-preserving model can be useful in cosmology. We present preliminary results in Appendix~\ref{app:preliminary} and we leave a more detailed investigation to future work.

\section{Conclusions}
\label{sec:conclusions}

\noindent
Linear methods such as singular value decomposition (SVD) are used for dimensionality reduction and data compression in various domains of physics and astronomy . In this paper, we have shown that autoencoders can be used to perform more optimal dimensionality reduction than linear methods in the context of gravitational wave data analysis. Specifically, we introduce a metric-preserving autoencoder whose latent space can be used to efficiently construct gravitational wave template banks and requires us to use just two dimensions. Comparing our new template banks with a bank constructed using two SVD dimensions, our banks perform much better in the effectualness test (see Figs.~\ref{fig:GWs_phase_overlap} and~\ref{fig:GWs_effectualness}). We also compare our results with the non-linear dimensionality-reduction method used in Ref.~\cite{Wadekar:2023kym} (where they predict higher-order SVD coefficients from lower-order ones with random forest regressors). We find that the effectualness of our template banks is comparable to that of \cite{Wadekar:2023kym} across the whole mass range, while our approach allows for using a slightly smaller number of templates.


Future directions might include extending our method to include higher modes as in \cite{Wadekar:2023kym} or effects of spin-orbit precession. We also plan to use the autoencoder-based template banks to perform a binary black hole merger search in the data from a LIGO-Virgo-Kagra (LVK) observing run. Other than building template banks, our autoencoder method can be useful for other GW applications requiring dimensionality reduction, such as gravitational waveform modeling \cite{Field:2011mf,Schmidt:2020yuu,Chua:2018woh,Tiglio:2021ysj,Purrer:2014fza,Canizares2015,Smith:2016qas,Field:2013cfa,Blackman2014,Herrmann2012}, fast parameter estimation \cite{Mes17_GstLAL_LowLatency}, and model-independent tests of general relativity \cite{Datta:2022izc, Saleem:2021nsb}.

\vskip 4pt

\noindent\textit{Acknowledgments ---} {\small We thank Mark Cheung and Lorenzo Bordin for useful discussions. GC acknowledges support from the Institute for Advanced Study where part of this work was carried out. We thank Thomas Bakx for providing the dataset ``\texttt{Plin\_CAMB\_LCDM\_Wide.dat}'' we used as training dataset for the \texttt{COBRA} autoencoder and for providing the code used in Ref.~\cite{Bakx:2024zgu} to get the SVD of the linear matter power spectra templates. We acknowledge use of the \href{https://numpy.org/}{\texttt{numpy}}, \href{https://scipy.org/}{\texttt{scipy}}, \href{https://pytorch.org/}{\texttt{pytorch}}, \href{https://zuko.readthedocs.io/stable/}{\texttt{zuko}}, \href{https://scikit-learn.org/}{\texttt{scikit-learn}} and \href{https://matplotlib.org/}{\texttt{matplotlib}} packages. DW is supported by NSF Grants No.~AST-2307146, No.~PHY-2513337, No.~PHY-090003, and No.~PHY-20043, by NASA Grant No.~21-ATP21-0010, by John Templeton Foundation Grant No.~62840, by the Simons Foundation [MPS-SIP-00001698, E.B.], by the Simons Foundation International [SFI-MPS-BH-00012593-02], and by Italian Ministry of Foreign Affairs and International Cooperation Grant No.~PGR01167.}

\section*{Appendix}

\setcounter{section}{0} 
\renewcommand{\thesubsection}{\Alph{subsection}} 
\setcounter{subsection}{0}

\noindent In the Appendix we discuss different results that were not of crucial importance for the main text, but that are nevertheless interesting to understand the geometry of our template banks. Appendix~\ref{app:preliminary} contains some preliminary results on the \texttt{COBRA} autoencoder used in the cosmology example in section~\ref{sec:Discussion:cosmology}.

\subsection{Toy example: embedding of a spherical shell} 
\label{app:sphere}

\begin{figure}[t]
\centering
\includegraphics[width = 0.495\textwidth]{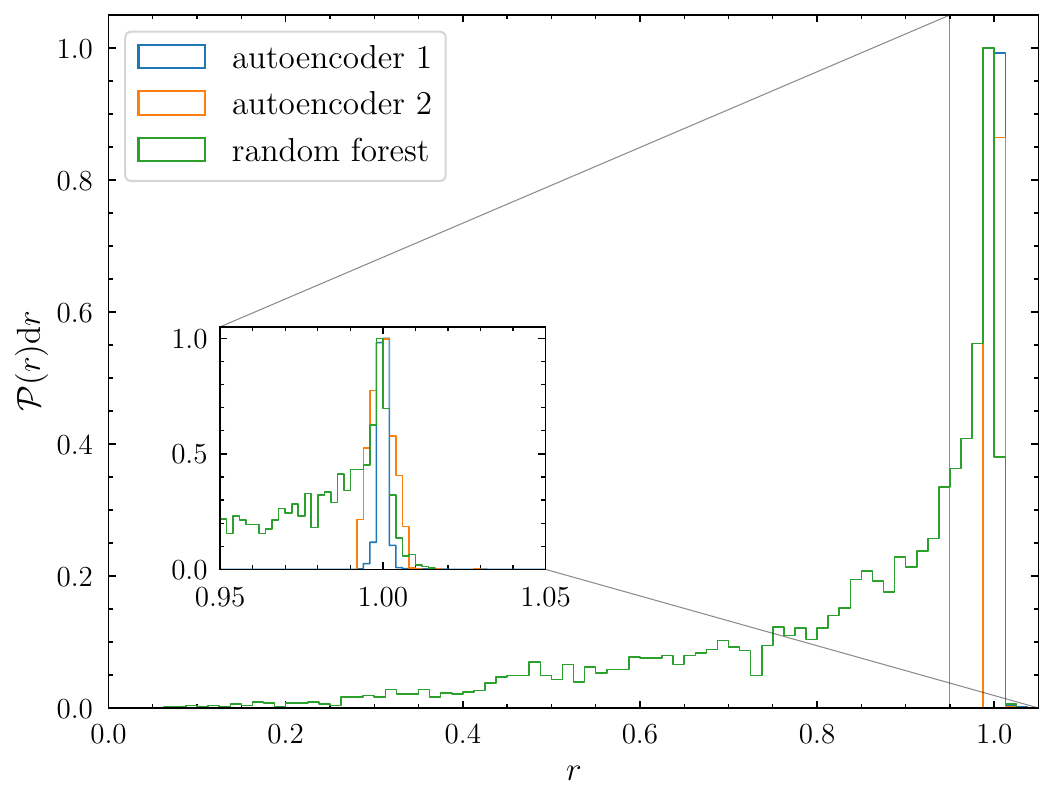} 
\includegraphics[width = 0.495\textwidth]{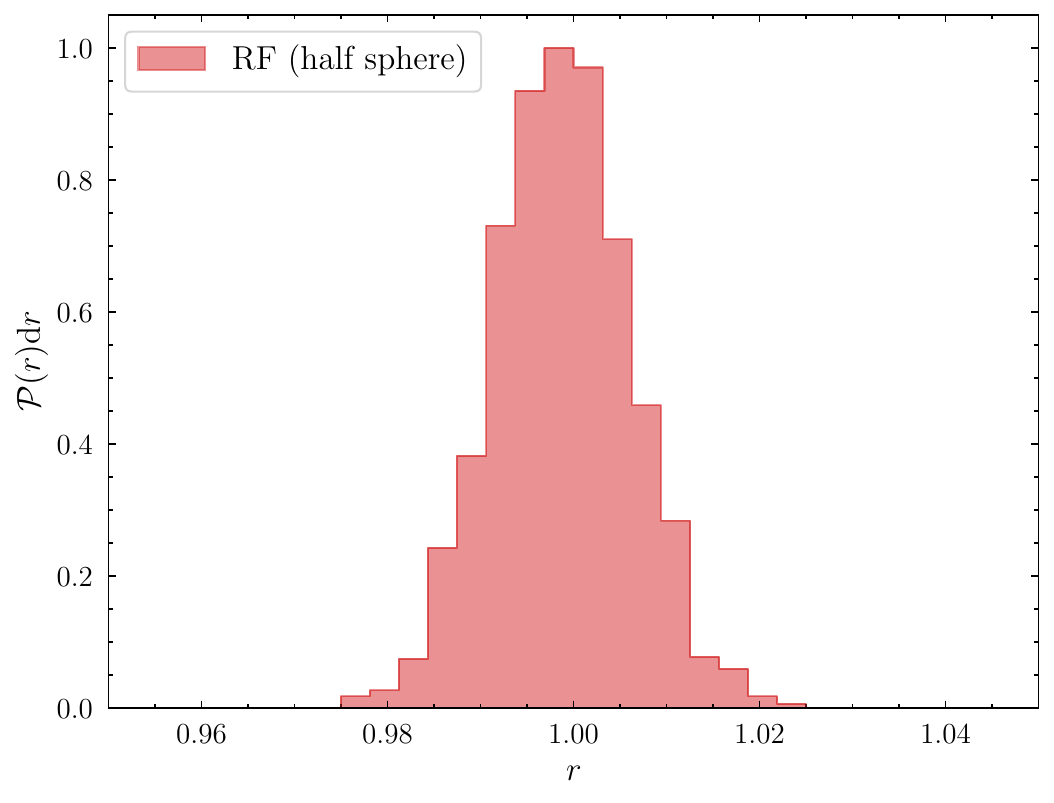} 
\caption{Left panel: histogram of validation radii $r=|{\bm x}|$ (where ${\bm x}=(x_{\rm valid},y_{\rm valid},{\rm RF}(x_{\rm valid},y_{\rm valid})$ for the random forest and ${\bm x} = {\cal D}({\cal E}({\bm x}_{\rm valid}))$ for the two autoencoders). Right panel: histogram of $r$ for the random forest in the case of the half sphere.}
\label{fig:sphere_autoencoder_vs_RF}
\end{figure}

\noindent It is always instructive to consider some examples that are well understood. We start with the unit sphere $S^2$ isometrically embedded in $3$-dimensional Euclidean space. We generate $8192$ points on the sphere using the Fibonacci sphere algorithm \cite{2010MatGe..42...49G}, and randomly split half of them in training and validation sets. We imagine that the coordinates ${\bm x}=(x,y,z)$ play the role of the first $10$ SVD coefficients $c^A$, $A=0,\dots 9$ for GW waveforms discussed in the introduction, and that we want to describe this $2$-dimensional manifold using dimensionality reduction techniques. It is clear that, in this particular example (where by construction we do not have any hierarchy between $x$, $y$ or $z$, since we only ``postulated'' that these come from a SVD), a dimensionality reduction consisting in taking only $2$ of the three ``SVD coefficients'' (i.e.~only $x$ and $y$, or only $y$ and $z$, etc.) would completely fail to capture the manifold, hence we immediately move to other methods. Inspired also here by Ref.~\cite{Wadekar:2023kym} and by the main text, we then train a random forest (also here we use \url{https://scikit-learn.org/stable/modules/generated/sklearn.ensemble.RandomForestRegressor.html}, with $50$ trees and maximum depth of $50$) to predict the $z$ coordinate as a function of $(x,y)$. We plot the result in the left panel of Fig.~\ref{fig:sphere_autoencoder_vs_RF}. We see that the random forest makes an extremely poor job of describing $S^2$. The reason is that for any given $(x,y,z)$ with unit norm, also $(x,y,{-z})$ has $|{\bm x}|=1$, and the random forest cannot capture this. This is confirmed by the right panel of Fig.~\ref{fig:sphere_autoencoder_vs_RF}, where we use a random forest with the same architecture to predict $z={\rm RF}(x,y)$ but only for $z\geq 0$: the random forest manages to fit the half sphere just fine.

We can then train an autoencoder $\smash{3 \overset{\raisebox{-1.5em}{${\scriptstyle \cal E}$}}{\longrightarrow} 2 \overset{\raisebox{-1.5em}{${\scriptstyle \cal D}$}}{\longrightarrow} 3}$ to learn a coordinate chart ${\bm x}={\bm x}(e^0,e^1)$ for $S^2$.\footnote{Let us recall that the sphere is not coverable by a single atlas. Even the standard $\theta,\phi$ coordinates fail at the poles (which are nevertheless sets of zero measure).} 
At variance with the random forest case, we now want to find a way to enforce that distances in the latent space are equal to distances in the embedding space (Euclidean distances $|\delta {\bm x}|$ between points on the unit sphere): this is because, as discussed in Section~\ref{sec:intro}, we are focusing on autoencoders that we can use in the GW analysis pipeline with geometric placement. 
Once we have the decoder ${\cal D}$, recalling the formula for the induced metric 
\begin{equation}
\label{eq:induced_metric}
    g_{\mu\nu}=\delta_{IJ}\frac{\partial{\cal D}^I}{\partial e^\mu}\frac{\partial{\cal D}^J}{\partial e^\nu}\,\,,
\end{equation}
we see that the requirement of distances in the latent space being equal to distances in the embedding space is equivalent to asking that $g_{\mu\nu}\approx\delta_{\mu\nu}$, and it can be imposed at two different levels:
\begin{enumerate}[leftmargin=*]
    \item \label{item:point_equality-1} we can ask that the equality holds around a point;
    \item \label{item:point_equality-2} we can ask that it holds globally on the sphere.
\end{enumerate}
It is always possible to satisfy \ref{item:point_equality-1} (we would just be looking for a Riemann coordinate system at that point), but it is never possible to satisfy \ref{item:point_equality-2} due to the curvature of the sphere (a Riemann patch will have size of at most $\sim 1/\sqrt{2}$, recalling that for $S^2$ the Ricci scalar $R$ is equal to $2$). 

\begin{figure}[t]
\centering
\includegraphics[width = 0.7\textwidth]{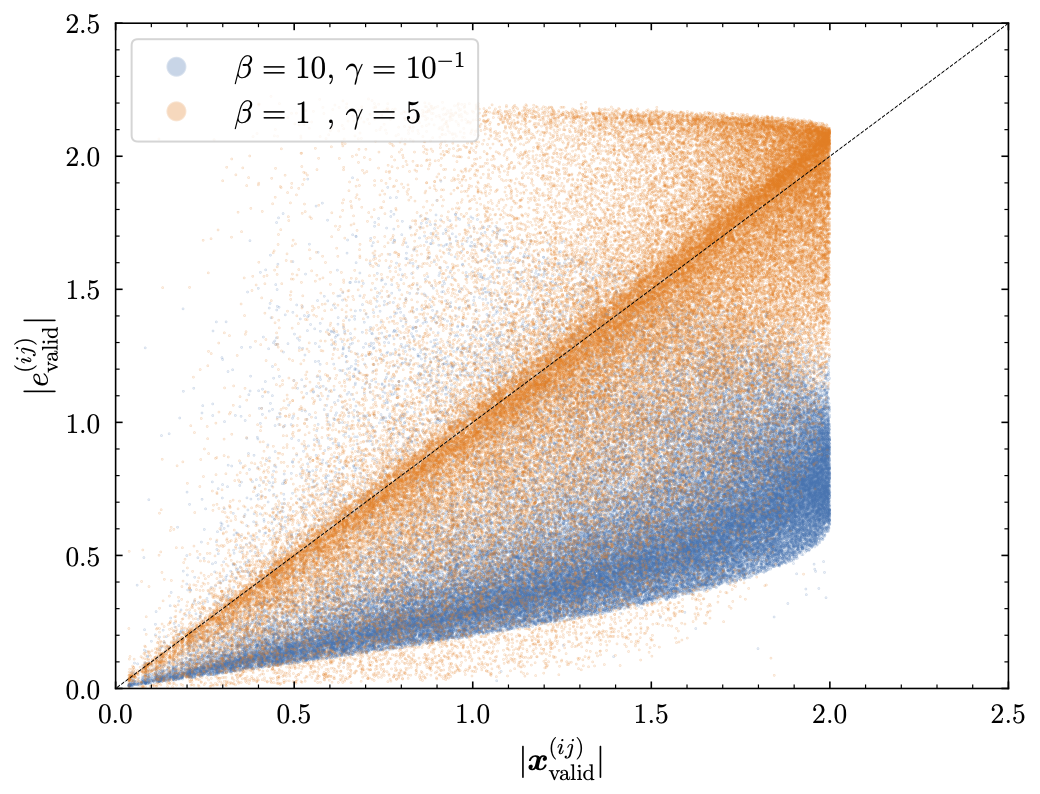} 
\caption{Same as Fig.~\ref{fig:GWs_autoencoder_distance_preservation_b_0_4_12} for the autoencoder describing the unit sphere, with two different choices of the loss function hyperparameters ($\alpha$ -- which is equal to $1$ in both cases -- $\beta$ and $\gamma$, see Eq.~\ref{eq:alpha_beta_loss}).}
\label{fig:autoencoder_distance_preservation}
\end{figure}

Nevertheless, let us try to train an autoencoder in \texttt{pytorch} that satisfies \ref{item:point_equality-2}, to see that this is not possible: as an example, we consider a simple architecture where $\cal E$ and $\cal D$ are both MLPs with $4$ hidden layers, each with $64$ neurons activated by the GELU function. As loss function we again use ($\smash{\tilde{x}^I_{\rm train} = {\cal D}^I({\cal E}({\bm x}_{\rm train}))}$ in the equation below)
\be
\label{eq:alpha_beta_loss}
\begin{split}
   L &= \alpha\times L_{\rm reconstruction} + \beta\times L_{\text{distance preservation}} \\
   &=\alpha\times{\rm MSE}({\bm x}_{\rm train},\tilde{\bm x}_{\rm train}) + \beta\times L_{\text{distance preservation}}\,\,.
\end{split}
\ee
Here the first term is a simple reconstruction loss for the autoencoder, while the distance-preservation loss is given by
\be
\label{eq:distance_preservation_loss_sphere}
L_{\text{distance preservation}} = \frac{2}{N_{\rm train}(N_{\rm train}-1)}\sum_{i<j} {\rm e}^{-\frac{|{\bm x}^{(ij)}_{\rm train}|}{\gamma}}\big(|{\bm x}^{(ij)}_{\rm train}| - |e^{(ij)}_{\rm train}|\big)^2\,\,,
\ee
where
\be
|{\bm x}^{(ij)}_{\rm train}| = |{\bm x}^{(i)}_{\rm train}-{\bm x}^{(j)}_{\rm train}|\qquad\text{and}\qquad|e^{(ij)}_{\rm train}| = \sqrt{\delta_{\mu\nu}\big[{\cal E}^\mu({\bm x}^{(i)}_{\rm train})-{\cal E}^\mu({\bm x}^{(j)}_{\rm train})\big]\big[{\cal E}^\nu({\bm x}^{(i)}_{\rm train})-{\cal E}^\nu({\bm x}^{(j)}_{\rm train})\big]}\,\,.
\ee
The hyperparameters $\alpha$ and $\beta$ control the relative strength of the two loss functions, while $\gamma$ as in Eq.~\eqref{eq:distance_preservation_loss_GWs} controls the ``locality'' of the distance preservation. Only distances of samples whose separation in the embedding space is of order $\gamma$ are preserved: so for small $\gamma$ we only preserve distances of nearby samples, while larger $\gamma$ imposes the constraints through a finite domain. The results are shown in Fig.~\ref{fig:autoencoder_distance_preservation} for two different (random) choices of $\alpha,\beta,\gamma$: we see that it is not possible to preserve distances globally on the sphere, as expected.\footnote{Since we know it is impossible to find a coordinate system on the sphere where the metric is flat, we did not investigate what happens if we try to change the architecture of the autoencoder.} In Appendix~\ref{app:rnc_autoencoder}, we show instead what happens if we try to satisfy \ref{item:point_equality-1}.

It is interesting to look at the curvature of the latent space in more detail, and see how to recover the curvature of the sphere. Since we use the $C^\infty$ GELU activation, we can use standard formulas of differential geometry and \texttt{pytorch}'s automatic differentiation to compute the metric of Eq.~\eqref{eq:induced_metric}. From that, we can compute the Christoffel symbols 
\be
\Gamma^\rho_{\mu\nu} = g^{\rho\lambda}\delta_{IJ}\frac{\partial{\cal D}^I}{\partial e^\lambda}\frac{\partial^2{\cal D}^J}{\partial e^\mu\partial e^\nu}\,\,,
\ee
and finally the Riemann tensor $\smash{R^\rho_{\hphantom{\rho}\sigma\mu\nu} = \partial_\mu\Gamma^\rho_{\nu\sigma}-\partial_\nu\Gamma^\rho_{\mu\sigma} + \Gamma^\rho_{\mu\lambda}\Gamma^\lambda_{\nu\sigma} - \Gamma^\rho_{\nu\lambda}\Gamma^\lambda_{\mu\sigma}}$. However, we have found it not so straightforward in \texttt{pytorch} to compute derivatives of a vector function -- the decoder, in our case -- past the second, which would be needed to compute the Riemann tensor. Hence we pursue a different route using the connection between holonomy and curvature, i.e.~the non-Abelian Stokes theorem -- see e.g. Appendix I of Ref.~\cite{Carroll:2004st} and Ref.~\cite{Broda:2000id}. Let us consider the parallel propagator $\smash{P^\mu_{\hphantom{\mu}\nu}(\lambda,\lambda_0)}$ along a path $e^\mu(\lambda)$, $\lambda\in[\lambda_0,\lambda_1]$. This satisfies the ODE
\be
\label{eq:parallel_propagator_definition}
\frac{{\rm d}}{{\rm d}\lambda}P^\mu_{\hphantom{\mu}\nu}(\lambda,\lambda_0) = A^\mu_{\hphantom{\mu}\sigma}(\lambda)P^\sigma_{\hphantom{\sigma}\nu}(\lambda,\lambda_0)\,\,,\quad\text{where}\quad A^\mu_{\hphantom{\mu}\sigma}(\lambda)={-\Gamma^\mu_{\rho\sigma}}\big|_{e=e(\lambda)}\frac{{\rm d}e^\rho(\lambda)}{{\rm d}\lambda}\quad\text{and}\quad P^\mu_{\hphantom{\mu}\nu}(\lambda_0,\lambda_0) = \delta^\mu_{\hphantom{\mu}\nu}\,\,.
\ee
In a $2$-dimensional manifold, if we consider a closed path $\partial {\cal S}_\epsilon$ and we compute the holonomy $\smash{P^\mu_{\hphantom{\mu}\nu}(\lambda_1,\lambda_0)\equiv P^\mu_{\hphantom{\mu}\nu}(\partial {\cal S}_\epsilon)}$ around that path, we have that
\be
\label{eq:holonomy_master_result_2D}
P^1_{\hphantom{1}0}(\partial {\cal S}_\epsilon) = \frac{1}{2}\int_{{\cal S}_\epsilon}{\rm d}^2e\,R(e)g_{00}(e) + \dots\,\,,
\ee
where $\dots$ denote higher orders in $\smash{\int_{{\cal S}_\epsilon}{\rm d}^2e}$. Hence we can define an estimator for the Ricci scalar at a point $e$ on the manifold by considering the holonomy around a circle $\partial D_\epsilon(e)$ of radius $\epsilon$ centered in $e$:
\be
\label{eq:ricci_estimator}
\frac{R(e)}{2}\approx\frac{P^1_{\hphantom{1}0}(\partial D_\epsilon(e))}{\int_{D_\epsilon(e)}{\rm d}^2e'\,g_{00}(e')}\,\,.
\ee

\begin{figure}[ht]
    \centering
    \settoheight{\FigHeight}{\includegraphics[width=0.99\textwidth]{plots_sphere/points_for_curvature_calculation.png}} 
    \begin{subfigure}{\textwidth}
        \centering
        \includegraphics[width=0.99\textwidth]{plots_sphere/points_for_curvature_calculation.png}
        \caption{\small In orange we plot the validation embeddings (right panel) and corresponding decoded coordinates in Euclidean space (left panel) for the autoencoder trained with $\alpha=1$, $\beta=1$ and $\gamma=5$. In black we show the $12288$ randomly-selected centers $e\in{\cal S}$ of disks $D_\epsilon(e)$ with radius $\epsilon=10^{-3}$ that we use to compute the curvature of the sphere using Eq.~\eqref{eq:ricci_estimator}.}
    \label{fig:points_for_curvature_calculation_sphere}
    \end{subfigure}
    \hfill
    \begin{subfigure}{\textwidth}
        \centering
        \resizebox{!}{\FigHeight}{\includegraphics[width=\textwidth]{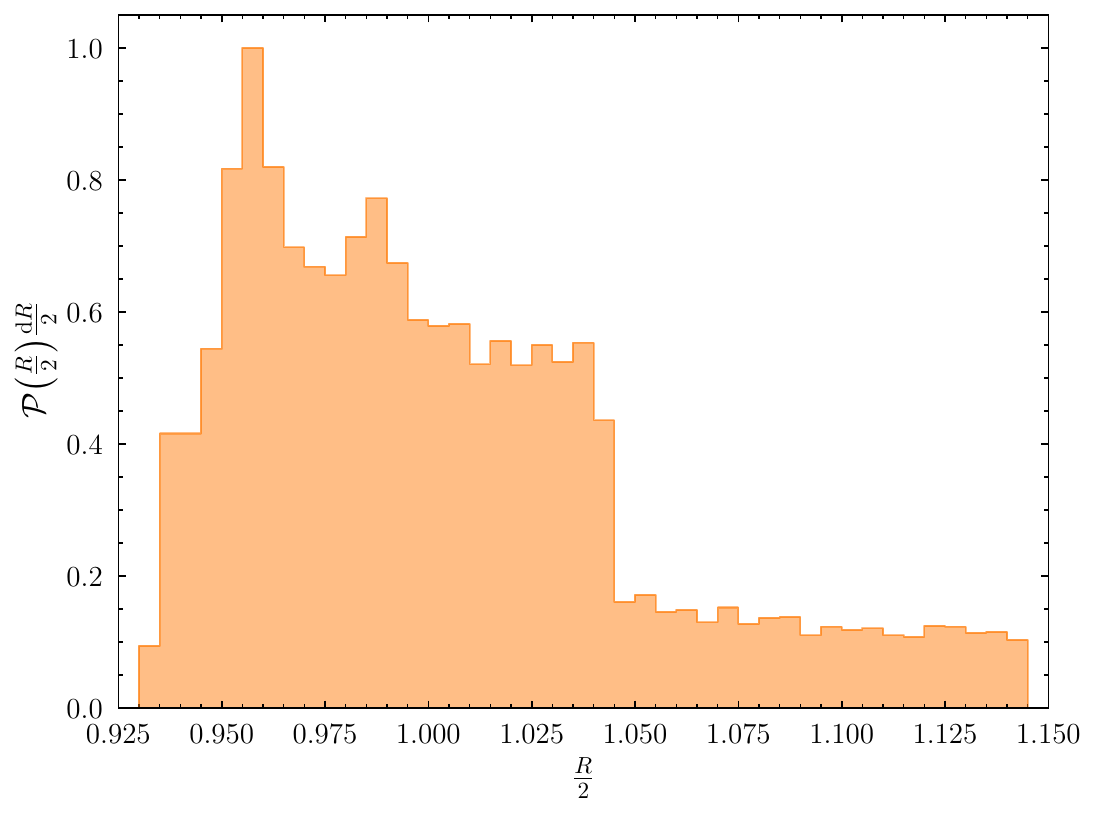}}
        \caption{\small Estimates, obtained using Eq.~\eqref{eq:ricci_estimator}, of the Ricci scalar in the region $\cal S$ shown in Fig.~\subfigref{\ref{fig:points_for_curvature_calculation_sphere_and_sphere_curvature}a}{fig:points_for_curvature_calculation_sphere}. We see that the distribution is consistent with the unit sphere having a constant Ricci curvature $R=2$.}
        \label{fig:sphere_curvature}
    \end{subfigure}
\invisiblecaption{fig:points_for_curvature_calculation_sphere_and_sphere_curvature}
\vspace*{-\baselineskip}
\end{figure}

In Fig.~\subfigref{\ref{fig:points_for_curvature_calculation_sphere_and_sphere_curvature}a}{fig:points_for_curvature_calculation_sphere} we show the validation embeddings and a collection of $12288$ randomly-chosen points $e\in{\cal S}$. This region is well inside the domain of the decoder (notice that for this experiment we consider only the autencoder trained with $\alpha=1$, $\beta=1$ and $\gamma=5$, shown in orange in Figs.~\ref{fig:sphere_autoencoder_vs_RF} and \ref{fig:autoencoder_distance_preservation}). Then, we compute the holonomy around circles of radius $\epsilon=10^{-3}$ centered at each of these points,\footnote{We solve the ODE of Eq.~\eqref{eq:parallel_propagator_definition} numerically, after refining the autoencoder to \texttt{torch.float64} precision.} and the integral inside the disk $D_\epsilon(e)$ that appears at the denominator of Eq.~\eqref{eq:ricci_estimator}. The result is shown in Fig.~\subfigref{\ref{fig:points_for_curvature_calculation_sphere_and_sphere_curvature}b}{fig:sphere_curvature}, where we see that the Ricci scalar is nicely distributed around the constant value $R=2$ as expected, consistently with errors coming from an inexact description of the sphere, higher orders in Eq.~\eqref{eq:holonomy_master_result_2D}, etc.

\begin{figure}[ht]
    \centering
    \settoheight{\FigHeight}{\includegraphics[width=0.99\textwidth]{plots_sphere/points_for_curvature_calculation.png}} 
    \begin{subfigure}{\textwidth}
        \centering
        \includegraphics[width=0.495\textwidth]{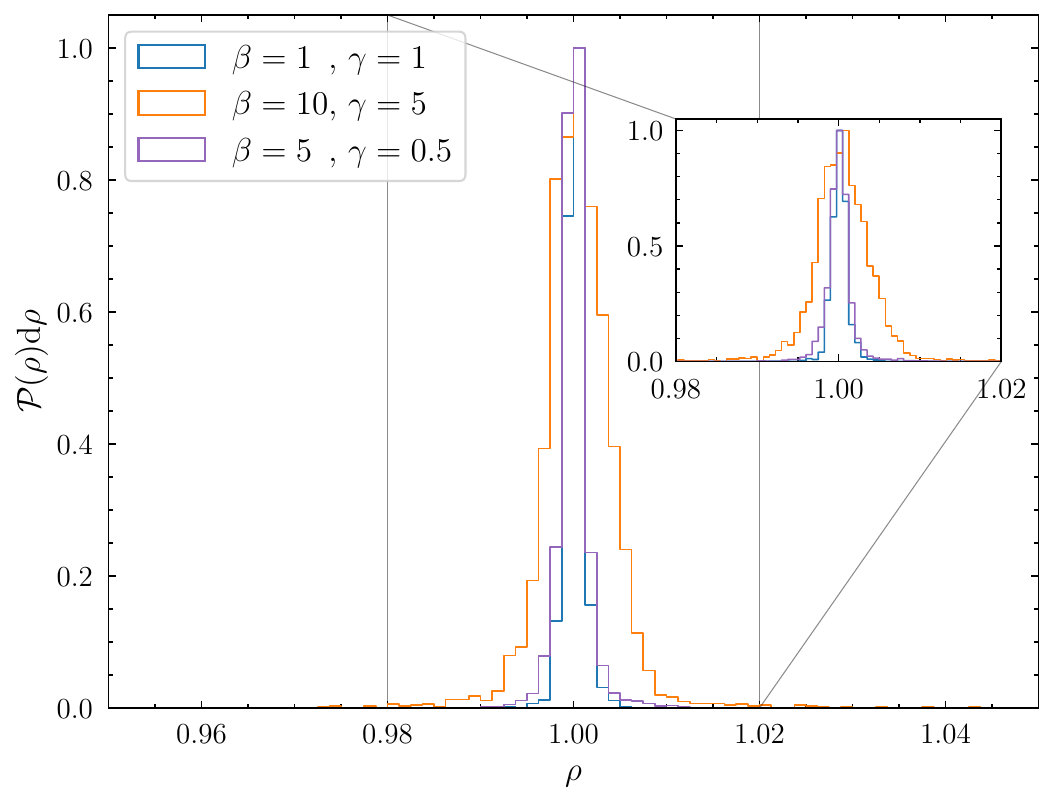}
        \includegraphics[width=0.495\textwidth]{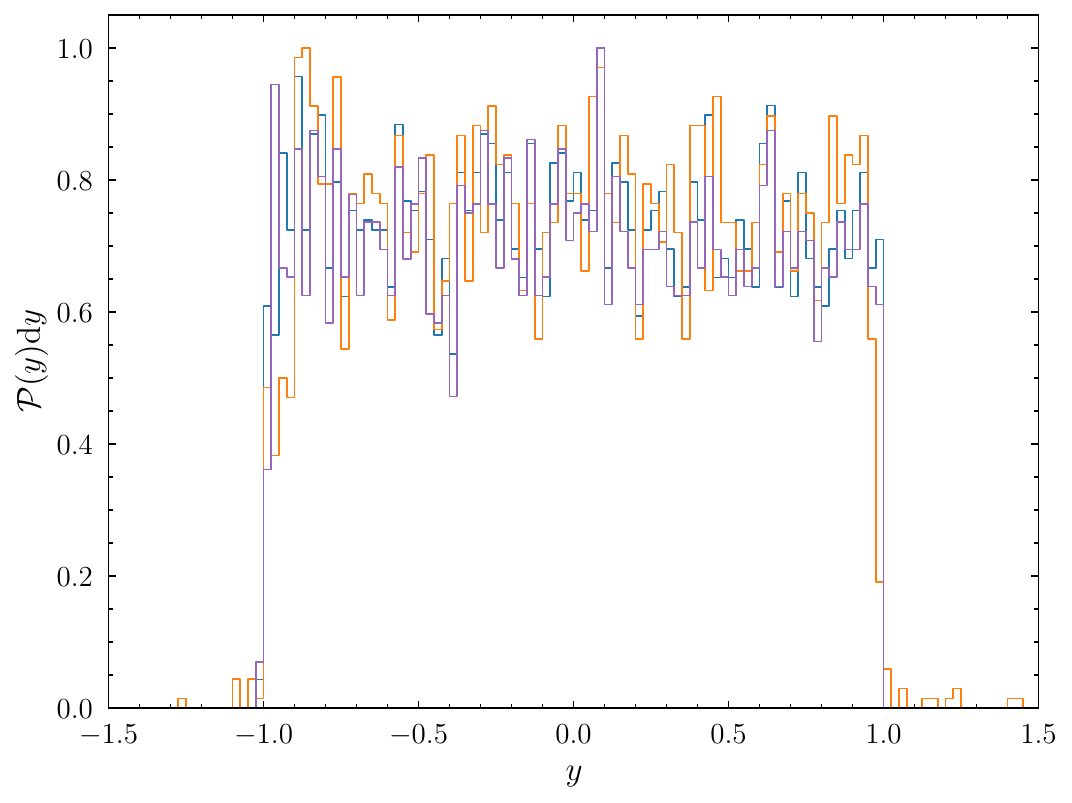}
        \caption{\small Similarly to Fig.~\ref{fig:sphere_autoencoder_vs_RF}, we show the histogram of validation radii ${\rho=\sqrt{x^2+z^2}}$ (left panel) and of validation $y$ coordinates (right panel) for the three autoencoders we trained to reconstruct the cylinder with the distance-preservation loss of Eq.~\eqref{eq:distance_preservation_loss_sphere}. In all three cases $\alpha$ is equal to $1$.}
    \label{fig:cylinder_reconstruction}
    \end{subfigure}
    \hfill
    \begin{subfigure}{\textwidth}
        \centering
        \resizebox{!}{\FigHeight}{\includegraphics[width=\textwidth]{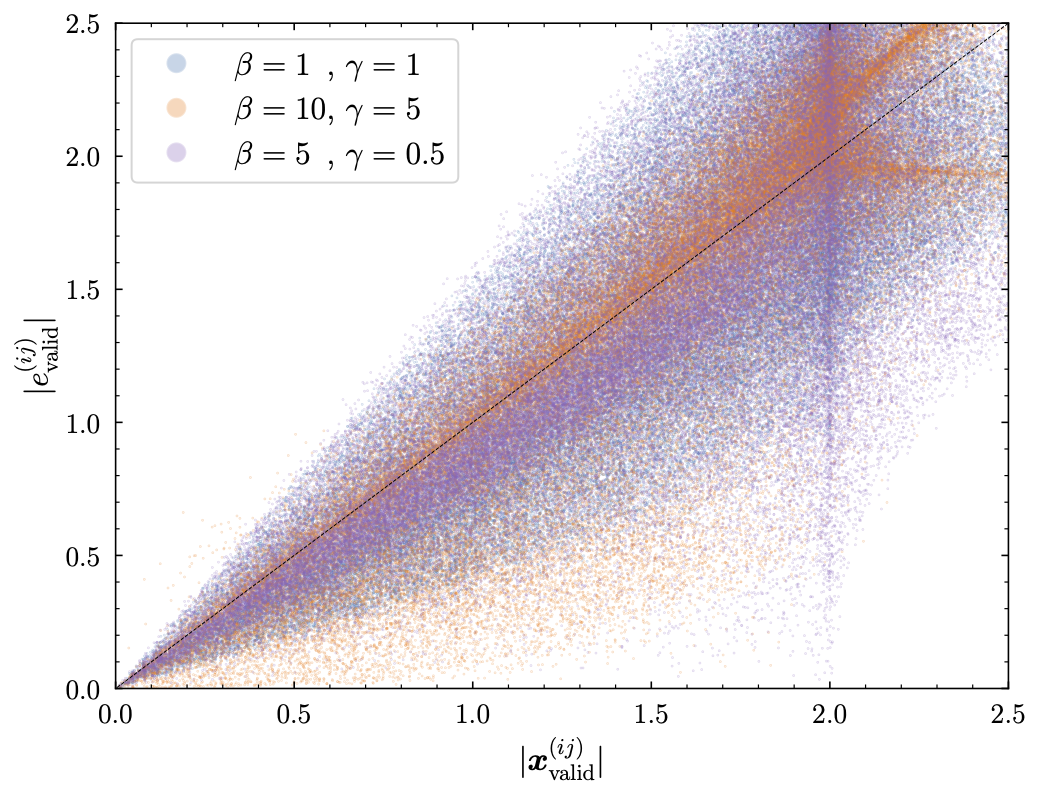}}
        \caption{\small The equivalent of Fig.~\ref{fig:autoencoder_distance_preservation} for the cylinder example.}
        \label{fig:cylinder_distance_preservation}
    \end{subfigure}
\invisiblecaption{fig:cylinder_main_text_figure}
\vspace*{-\baselineskip}
\end{figure}

\subsection{Toy example: embedding of a cylindrical shell} 
\label{app:cylinder}

We can then discuss another example, that of a cylinder. Contrarily to the sphere this is a flat surface, so we expect to be able to find a metric-preserving autoencoder. We generate $8192$ random points on a cylinder $\smash{\{x^2+z^2=1, -1\leq y\leq 1\}}$\footnote{We can easily generate an uniform set of points by randomly sampling $\theta$ and $y$, since the metric is flat in these coordinates.} and, using half these points as training and half as validation, we train six different autoencoders: three with the same architecture as the sphere, and three with $8$ hidden layers each with $128$ neurons. Each triple was trained with the loss of Eqs.~\eqref{eq:alpha_beta_loss},~\eqref{eq:distance_preservation_loss_sphere}: all have $\alpha=1$, while $\beta$ and $\gamma$ are respectively $\beta=1,\gamma=1$, $\beta=10,\gamma=5$, and finally $\beta=5,\gamma=0.5$. The results for the two architectures are similar, so in Fig.~\ref{fig:cylinder_main_text_figure} we show only the autoencoders with the same architecture as the sphere ones. While the radius is nicely fixed to $1$ and $y$ is restricted between ${-1}$ and $1$ (Fig.~\subfigref{\ref{fig:cylinder_main_text_figure}a}{fig:cylinder_reconstruction}), we see that distances are not preserved even for points that are close in the embedding space (Fig.~\subfigref{\ref{fig:cylinder_main_text_figure}b}{fig:cylinder_distance_preservation}). We suspect that this has to do with the fact that the cylinder, while flat, is closed and without boundary in the $\theta$ dimension. We did not investigate in detail if this is why the distance preservation of Eqs.~\eqref{eq:alpha_beta_loss},~\eqref{eq:distance_preservation_loss_sphere} fails (e.g.~by trying with a half cylinder, $\theta\in[0,\pi]$), also because in the main text we have seen that in the case of GW template banks the equivalent of the loss of Eq.~\eqref{eq:distance_preservation_loss_sphere}, adapted to a $10$-dimensional embedding space, is sufficient to obtain distance-preserving autoencoders. In Appendix~\ref{app:cylinder_nice}, however, we pursue a different strategy. Since the cylinder is flat, we know that we can have Riemann normal coordinates simultaneously around \emph{every} point on the cylinder: hence, we implement the condition \ref{item:point_equality-1} on all points in the training set instead of only around a point, and we show that this allows us to find a coordinate system where the metric is $\delta_{\mu\nu}$ and distances are preserved.

\subsection{\texorpdfstring{``Riemann normal coordinates'' autoencoder loss for unit sphere}{"Riemann normal coordinates" autoencoder loss for unit sphere}}
\label{app:rnc_autoencoder}

\begin{figure}[ht]
    \centering
    \settoheight{\FigHeight}{\includegraphics[width=0.99\textwidth]{plots_sphere/points_for_curvature_calculation.png}}
    \begin{subfigure}{\textwidth}
        \centering
        \includegraphics[width = 0.495\textwidth]{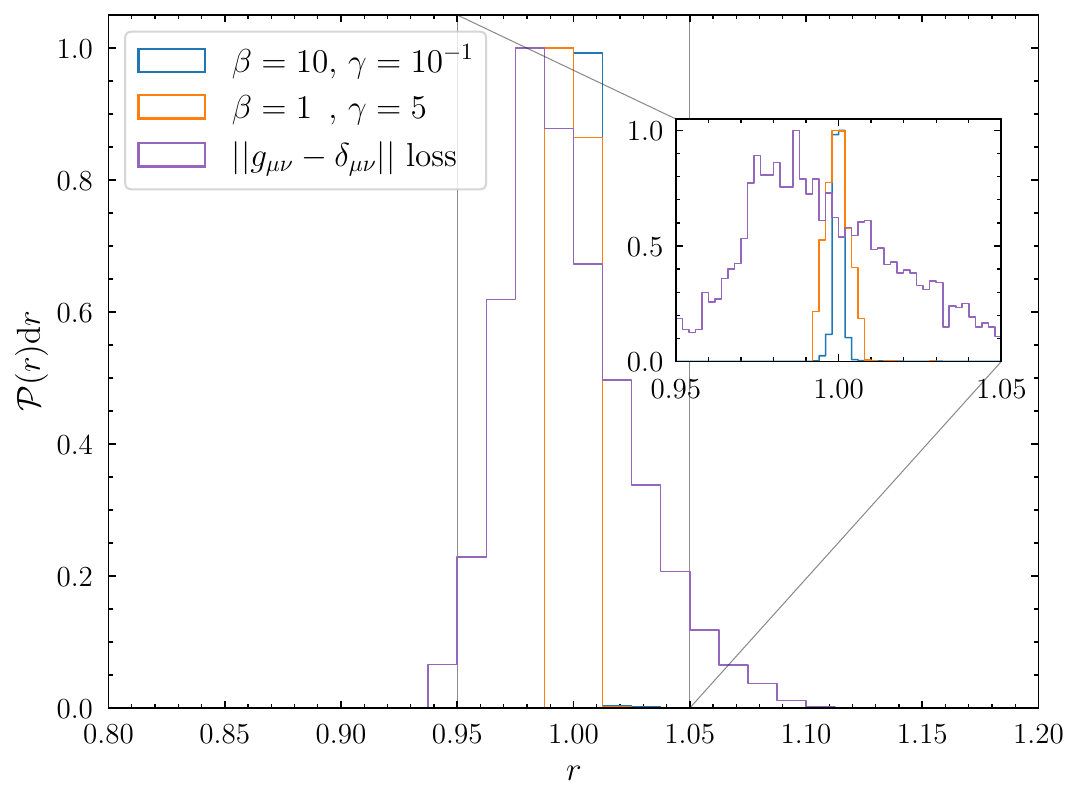} 
        \includegraphics[width = 0.495\textwidth]{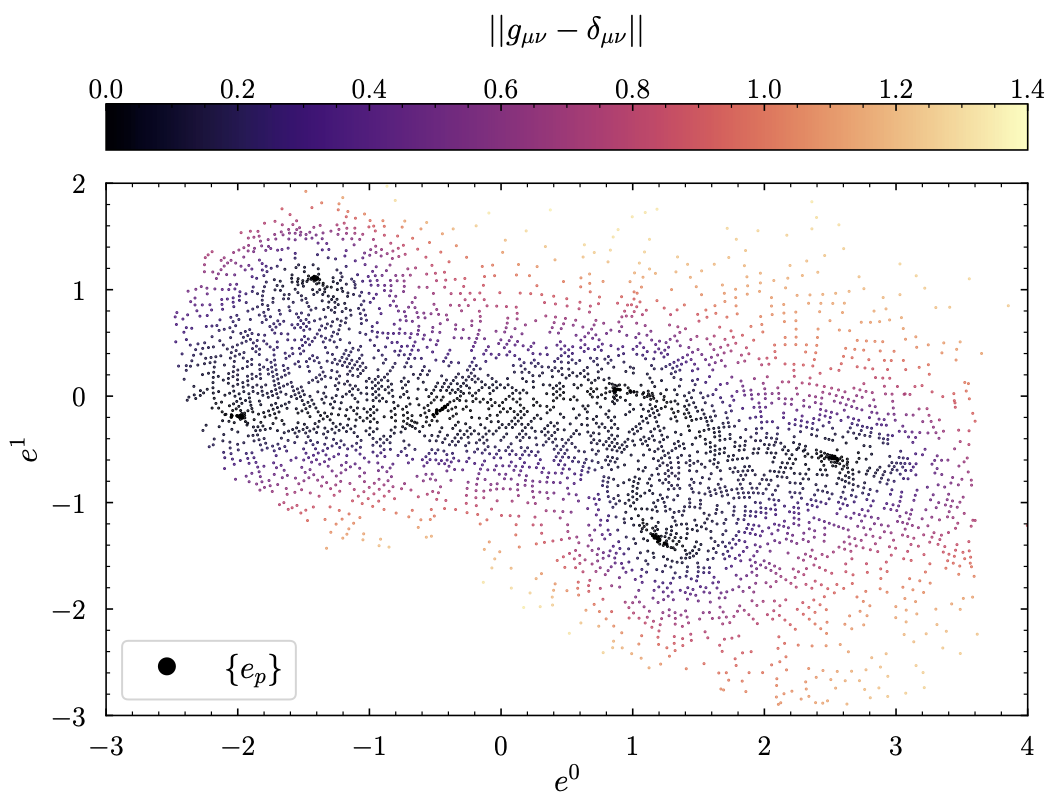} 
        \caption{\small Left panel: as in Fig.~\ref{fig:sphere_autoencoder_vs_RF}, we show the histogram of validation radii $r=|{\bm x}|$ for the two autoencoders of Appendix~\ref{app:sphere} and the new autoencoder trained with Eq.~\eqref{eq:fancy_loss}. Right panel: we plot of the difference between the induced metric and the identity matrix for the new autoencoder and the points $\{e_p\}$ where we try to find a Riemann coordinate system.}
        \label{fig:sphere_autoencoder_old_vs_new}
    \end{subfigure}
    \hfill
    \begin{subfigure}{\textwidth}
        \centering
        \resizebox{!}{\FigHeight}{\includegraphics[width=\textwidth]{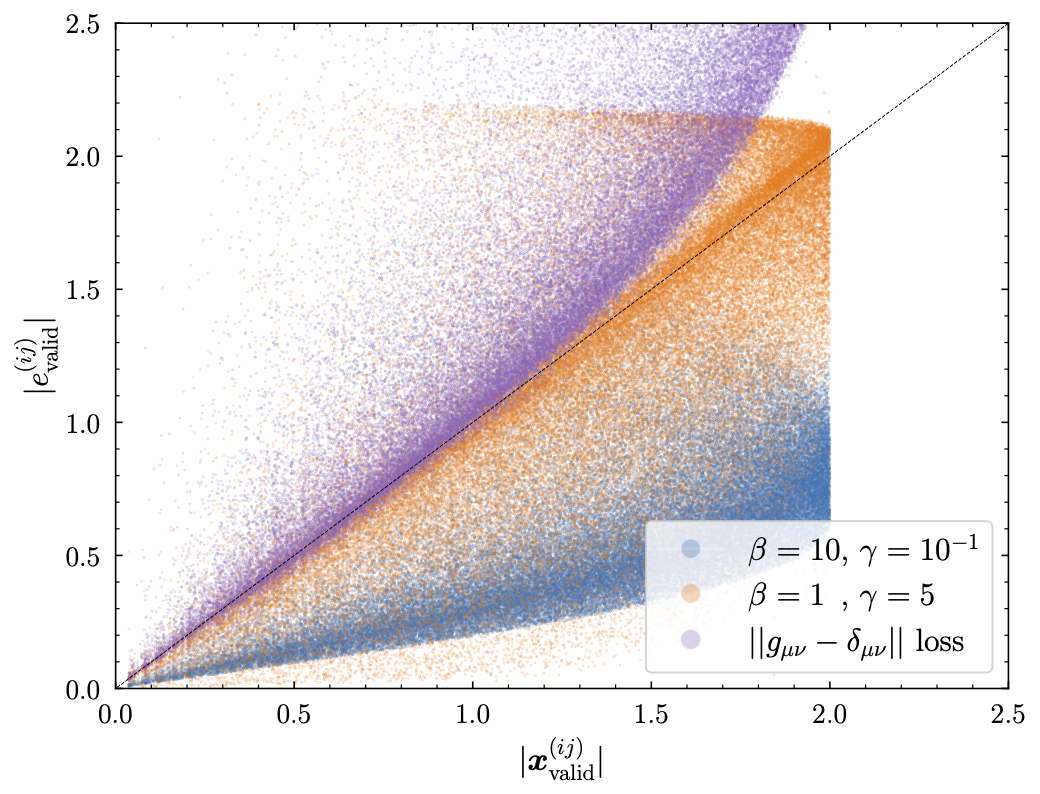}}
        \caption{\small Scatter plot of the distances in the embedding space vs.~the Euclidean distances in the latent space for $1\%$ of pairs of validation points, for the autoencoder trained with the loss of Eq.~\eqref{eq:fancy_loss} in addition to the two autoencoders shown in Fig.~\ref{fig:autoencoder_distance_preservation}.}
        \label{fig:autoencoder_distance_preservation.appendix}
    \end{subfigure}
\invisiblecaption{fig:sphere_autoencoder_old_vs_new_and_autoencoder_distance_preservation.appendix}
\vspace*{-\baselineskip}
\end{figure}

\noindent In this appendix, we show the results of training the same autoencoder of Appendix~\ref{app:sphere} with a distance preservation term in the loss different than Eq.~\eqref{eq:distance_preservation_loss_sphere}. More precisely, the loss we use now is
\be
\label{eq:fancy_loss}
L_{\text{distance preservation}} = \frac{1}{n_{\rm train}}\sum_{p=1}^{n_{\rm train}} ||g_{\mu\nu}(e_p)-\delta_{\mu\nu}||\,\,,
\ee
where $e_p = {\cal E}({\bm x}_p)$ are the latent coordinates of the $n_{\rm train}$ ${\bm x}_p$ points in the training set that are the closest (with distance within $5\times10^{-3}$) to the unit cube enclosing the unit sphere.\footnote{Notice that when training with this loss we put all the training data in a single batch.} Here $||\cdot||$ is the Frobenius norm (\emph{not} the Frobenius norm squared) and $g_{\mu\nu}$ is the metric of Eq.~\eqref{eq:induced_metric}. We also fix $\alpha=1=\beta$. The results are shown in Fig.~\subfigref{\ref{fig:sphere_autoencoder_old_vs_new_and_autoencoder_distance_preservation.appendix}a}{fig:sphere_autoencoder_old_vs_new}: we see that the autoencoder performs sensibly worse in reconstructing the sphere than those discussed in the main text. It is important to emphasize, though, that we did not try to refine the training of the autoencoder in any way after a first and relatively rough pass.

Fig.~\subfigref{\ref{fig:sphere_autoencoder_old_vs_new_and_autoencoder_distance_preservation.appendix}b}{fig:autoencoder_distance_preservation.appendix} is the equivalent of Fig.~\ref{fig:autoencoder_distance_preservation}, including also the autoencoder trained with the new loss. While we see that even with this new loss the distances are not preserved as well as for the template banks discussed in Section~\ref{subsec:autoencoders_GWs}, there is a marked improvement over the autoencoders of Appendix~\ref{app:sphere}, especially for small distances. It is also interesting that in this case all points seem to lie above the diagonal: while distances are not preserved, the latent-space ones seem to be consistently larger than the embedding-space ones. It would be interesting to investigate this example in more detail, possibly changing Eq.~\eqref{eq:fancy_loss} to compute the Frobenius loss squared, or choosing different values for the $\alpha/\beta$ ratio.

\subsection{\texorpdfstring{``Riemann normal coordinates'' autoencoder loss for cylinder}{"Riemann normal coordinates" autoencoder loss for cylinder}}
\label{app:cylinder_nice}

\begin{figure}[ht]
    \centering
    \settoheight{\FigHeight}{\includegraphics[width=0.99\textwidth]{plots_sphere/points_for_curvature_calculation.png}}
    \begin{subfigure}{\textwidth}
        \centering
        \includegraphics[width = 0.495\textwidth]{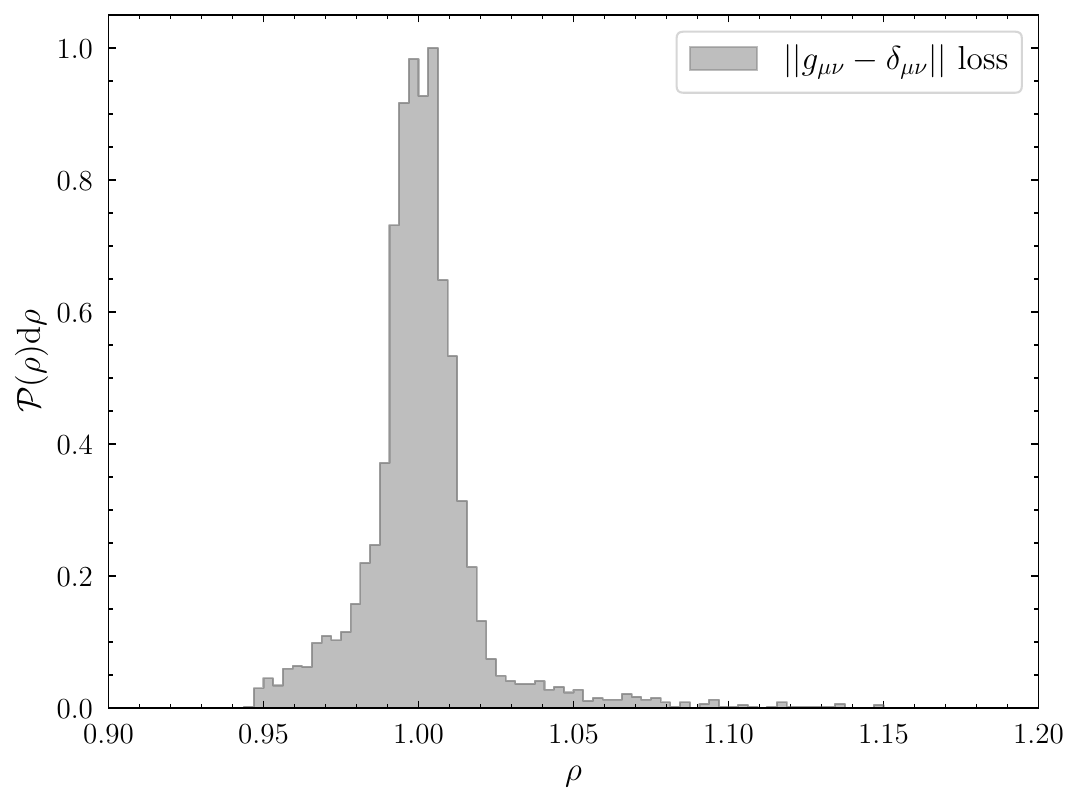} 
        \includegraphics[width = 0.495\textwidth]{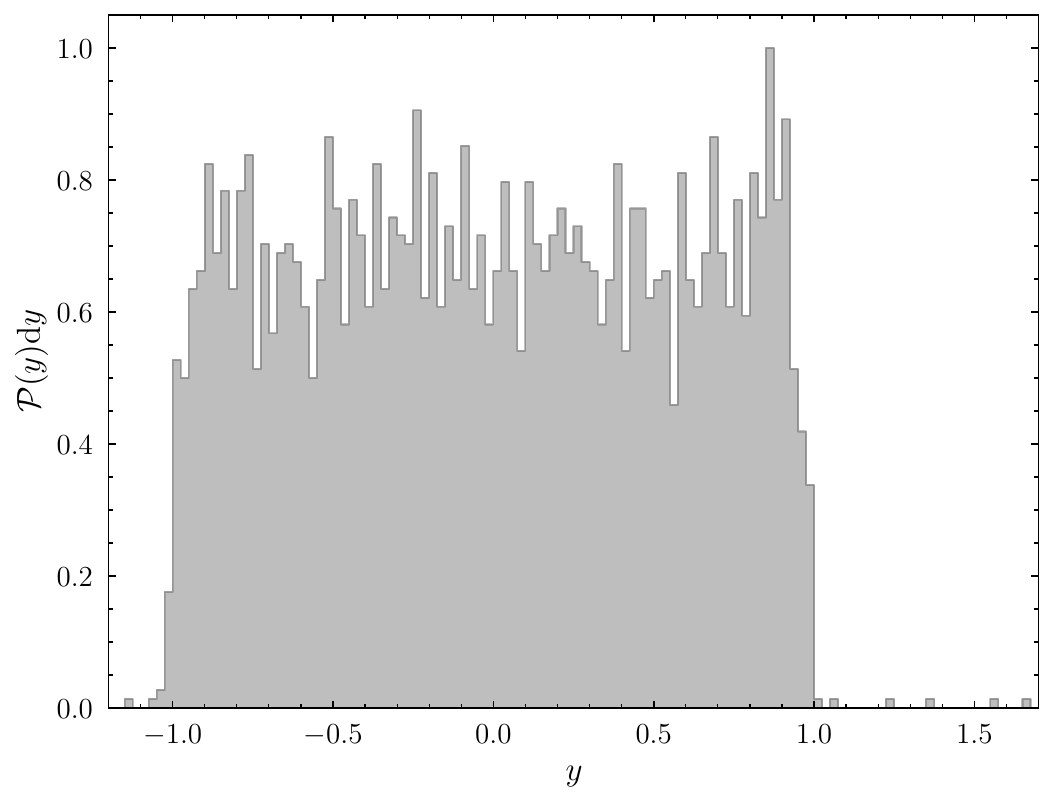} 
        \caption{\small Same as in Fig.~\subfigref{\ref{fig:cylinder_main_text_figure}a}{fig:cylinder_reconstruction}, but for the autoencoder trained with the loss of Eq.~\eqref{eq:fancy_loss} (imposed over all the training set in the case of a flat manifold like the cylinder).}
        \label{fig:cylinder_appendix-1}
    \end{subfigure}
    \hfill
    \begin{subfigure}{\textwidth}
        \centering
        \includegraphics[width = 0.495\textwidth]{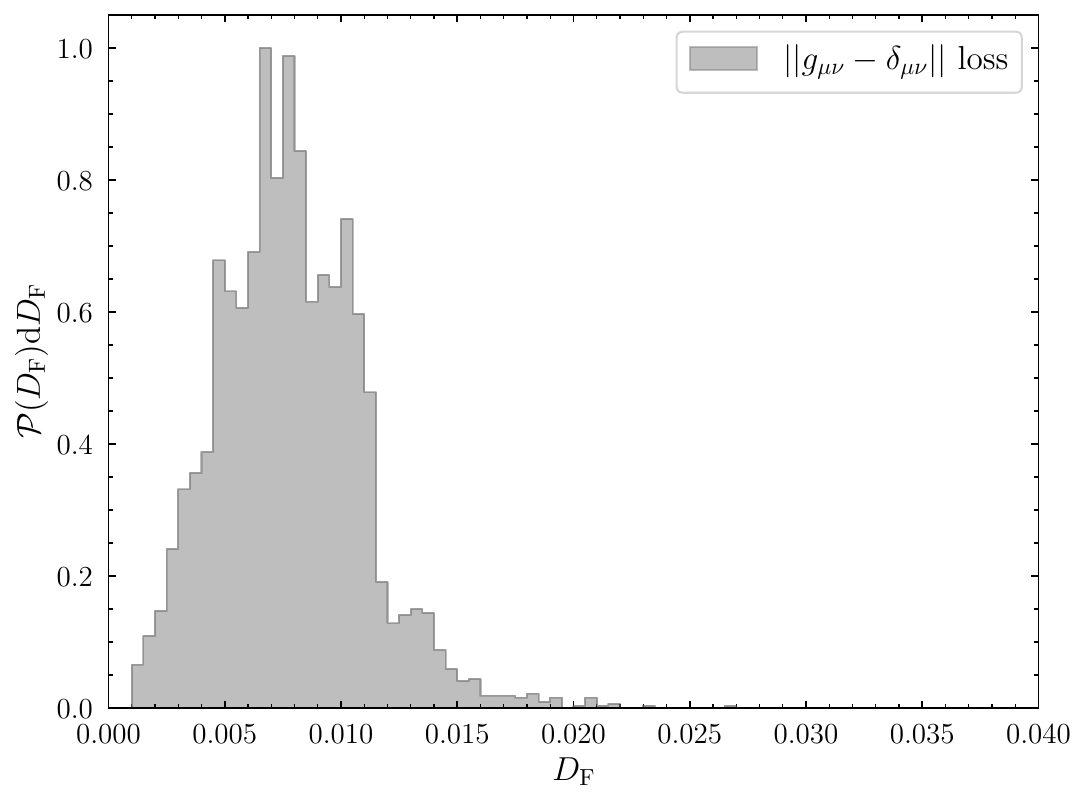} 
        \includegraphics[width = 0.495\textwidth]{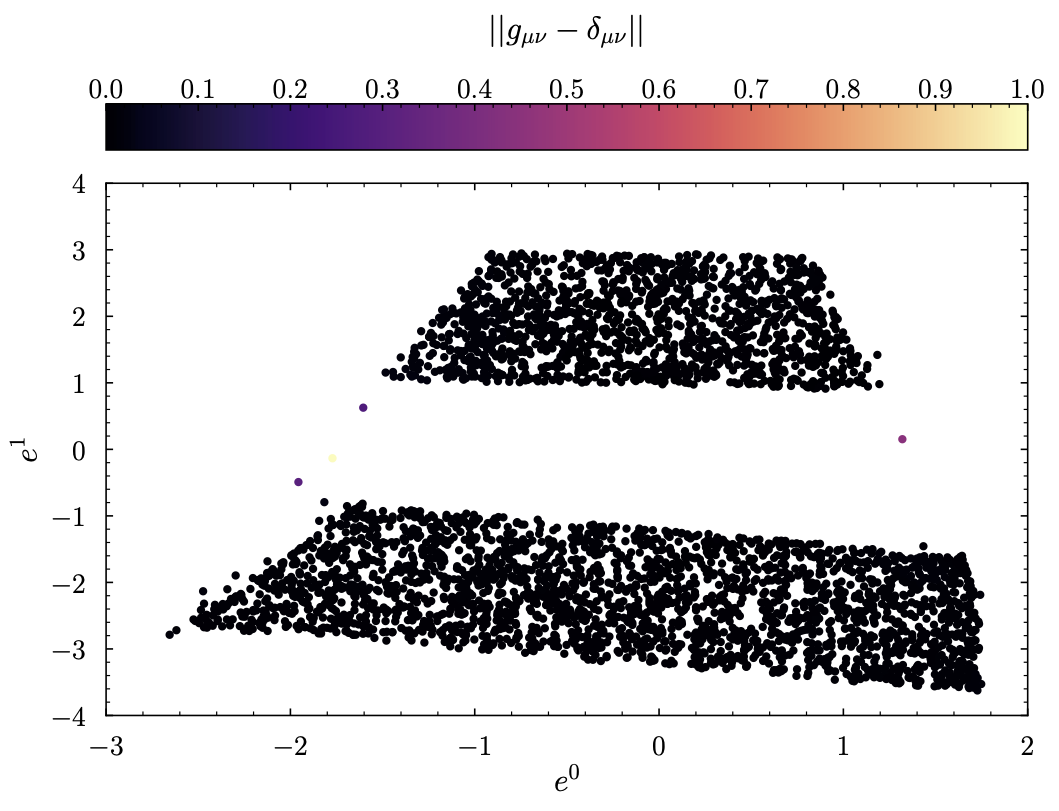}
        \caption{\small The left panel shows the distribution of validation Frobenius distances $\smash{||g_{\mu\nu}(e_{\rm valid})-\delta_{\mu\nu}||=D_{\rm F}}$ for the autoencoder of Fig.~\subfigref{\ref{fig:cylinder_appendix}a}{fig:cylinder_appendix-2}: only $4$ points have $D_{\rm F}>0.04$. The right panel shows a scatter plot similar to the right panel of Fig.~\subfigref{\ref{fig:sphere_autoencoder_old_vs_new_and_autoencoder_distance_preservation.appendix}a}{fig:sphere_autoencoder_old_vs_new}: we show all validation latent-space points colored by the value of $D_{\rm F}$, confirming the fact that the loss of Eq.~\eqref{eq:fancy_loss} succeeds at finding a coordinate system where the metric is flat.}
        \label{fig:cylinder_appendix-2}
    \end{subfigure}
\invisiblecaption{fig:cylinder_appendix}
\vspace*{-\baselineskip}
\end{figure}

\begin{figure}[t]
\centering
\includegraphics[width = 0.7\textwidth]{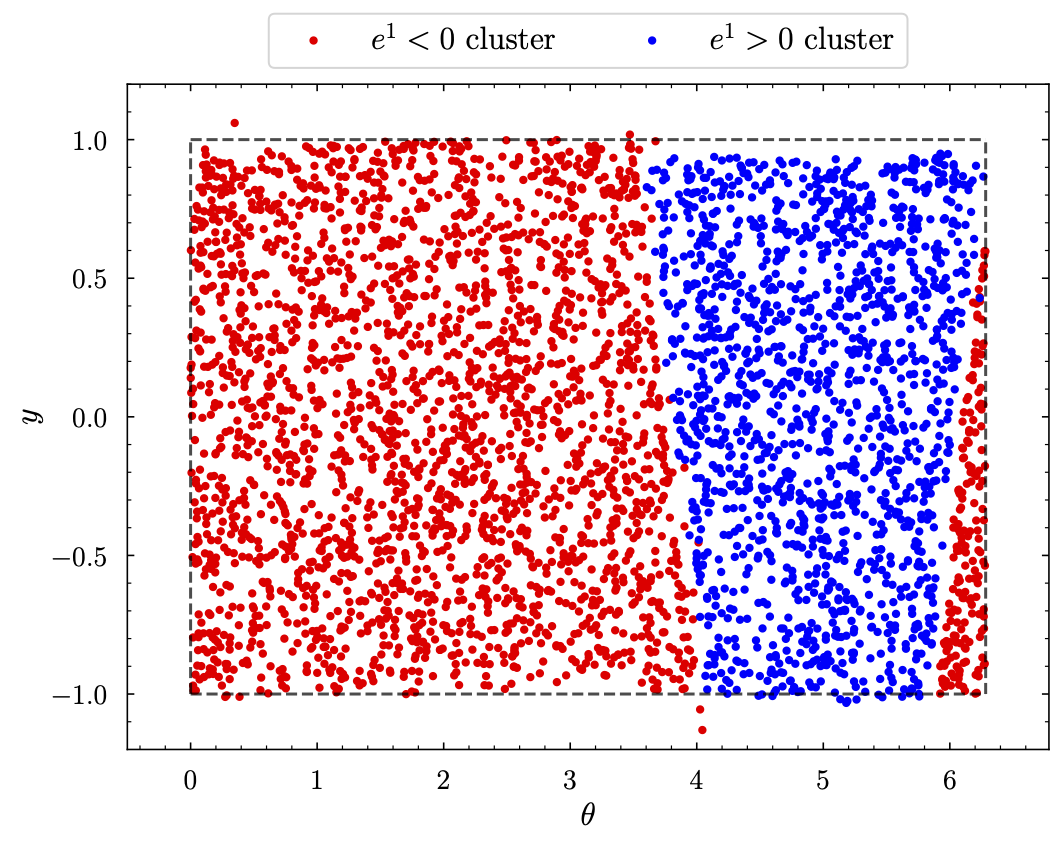} 
\caption{The two regions in $(\theta,y)$ space corresponding to the two clusters shown in the right panel of Fig.~\subfigref{\ref{fig:cylinder_appendix}b}{fig:cylinder_appendix-2}. We drop the $4$ points with $D_{\rm F}>0.04$ that -- not coincidentally -- are the ones that in that figure do not fall in either of the clusters but instead ``link'' the two clusters.}
\label{fig:cylinder_halves}
\end{figure}

\noindent In Appendix~\ref{app:cylinder} we showed that using the loss of Eq.~\eqref{eq:distance_preservation_loss_sphere} we cannot find a coordinate system on the cylinder where the metric is flat. Let us then investigate a different loss, similar to Eq.~\eqref{eq:fancy_loss}. Unlike in Appendix~\ref{app:rnc_autoencoder}, we take the ${\bm x}_p$ points to be all the training points, not just a subset of them (so now we have $n_{\rm train}=N_{\rm train}$). We train an autoencoder with the same architecture as in Appendix~\ref{app:sphere} with this loss, and show the results in Fig.~\ref{fig:cylinder_appendix}.

Besides the fact that we manage to find a coordinate system where the metric is flat using this loss, another interesting takeaway seems to be that the the autoencoder naturally seems to split the cylinder in two ``halves'', as we see in the right panel of Fig.~\subfigref{\ref{fig:cylinder_appendix}b}{fig:cylinder_appendix-2}. What the two halves correspond to in terms of $(\theta,y)$ variables is shown in Fig.~\ref{fig:cylinder_halves}. It would be interesting to investigate if conditional autoencoders (see e.g.~\url{https://behavenet.readthedocs.io/en/develop/source/user_guide.conditional_autoencoders.html}) could be employed for our GW template banks if we were to encounter the case where the manifold is flat but without boundary, possibly allowing us to keep the simple loss of Eq.~\eqref{eq:distance_preservation_loss_GWs} and still finding a set of multiple coordinate systems where the metric is close to $\smash{{
\cal N}^2\delta_{\mu\nu}}$. We also notice that the reconstruction of $\smash{\rho=\sqrt{x^2+z^2}}$ and $y$ is not as good as the autoencoders in Appendix~\ref{app:cylinder}: this could possibly be improved with a change of architecture and also by not imposing the distance preservation loss at every training point, but only over a subset of them. More precisely, one should batch the training set, and impose the distance preservation loss over a random subset of each batch at every training epoch.\footnote{Notice that all the results for the cylinder presented in this appendix and in Appendices~\ref{app:sphere} and \ref{app:cylinder} were obtained without batching the training dataset. Regarding the results in the main text, we did not investigate if batching could help with obtaining a better distance preservation than the one shown in Fig.~\subfigref{\ref{fig:cylinder_main_text_figure}b}{fig:cylinder_distance_preservation}. Given how bad the loss of Eq.~\eqref{eq:distance_preservation_loss_sphere} fails at finding a coordinate system where $\smash{g_{\mu\nu}\approx\delta_{\mu\nu}}$, even with varying $\alpha/\beta$ ratio and $\gamma$, we do not expect this would change Fig.~\subfigref{\ref{fig:cylinder_main_text_figure}b}{fig:cylinder_distance_preservation} much.}

\subsection{Curvature of GW template banks}
\label{app:autoencoder_curvature}

\begin{figure}[t]
\centering
\includegraphics[width = 0.7\textwidth]{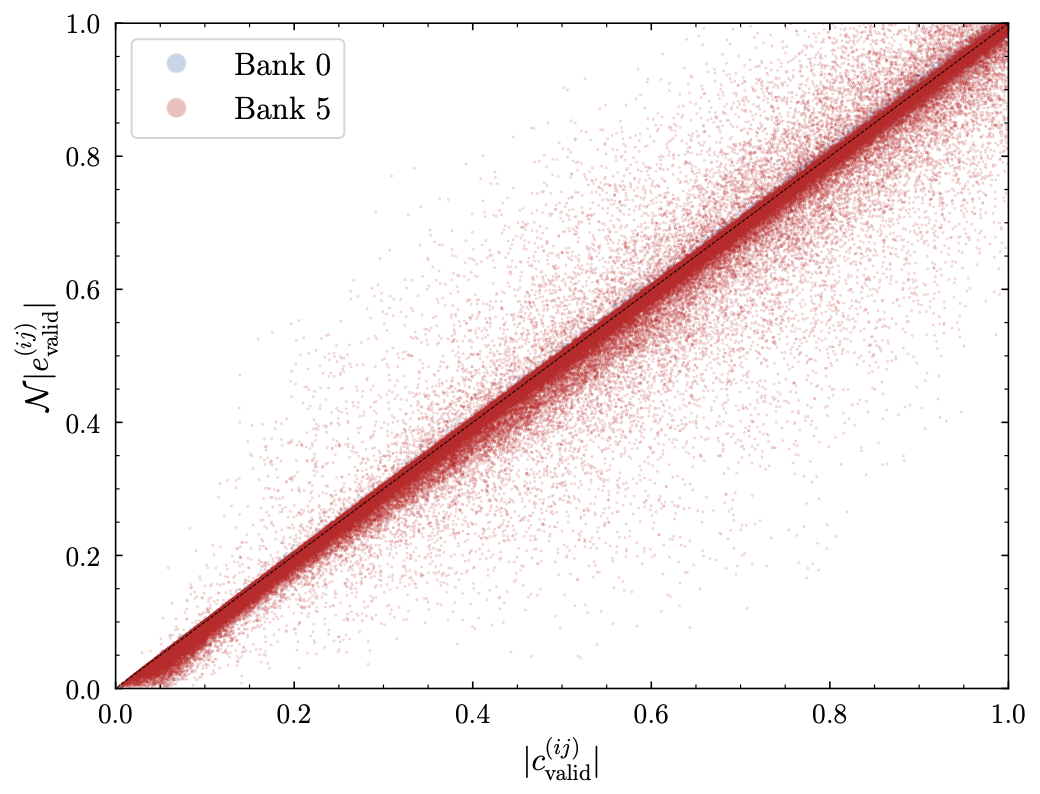} 
\caption{Same as Fig.~\ref{fig:GWs_autoencoder_distance_preservation_b_0_4_12}, but for Bank $0$ and the bank $\text{BBH-5}, {\cal M}\in[14.7,23.9]\,{M_\odot}$ (Bank $5$). At variance with Fig.~\ref{fig:GWs_autoencoder_distance_preservation_b_0_4_12} we plot all pairs of distances such that $\smash{|{c}^{(ij)}_{\rm valid}|\leq 1}$, to emphasize how for Bank $5$ distances are not preserved as well as the others.}
\label{fig:GWs_autoencoder_distance_preservation_b_0_5}
\end{figure}

\begin{figure}[t]
\centering
\includegraphics[width = 0.495\textwidth]{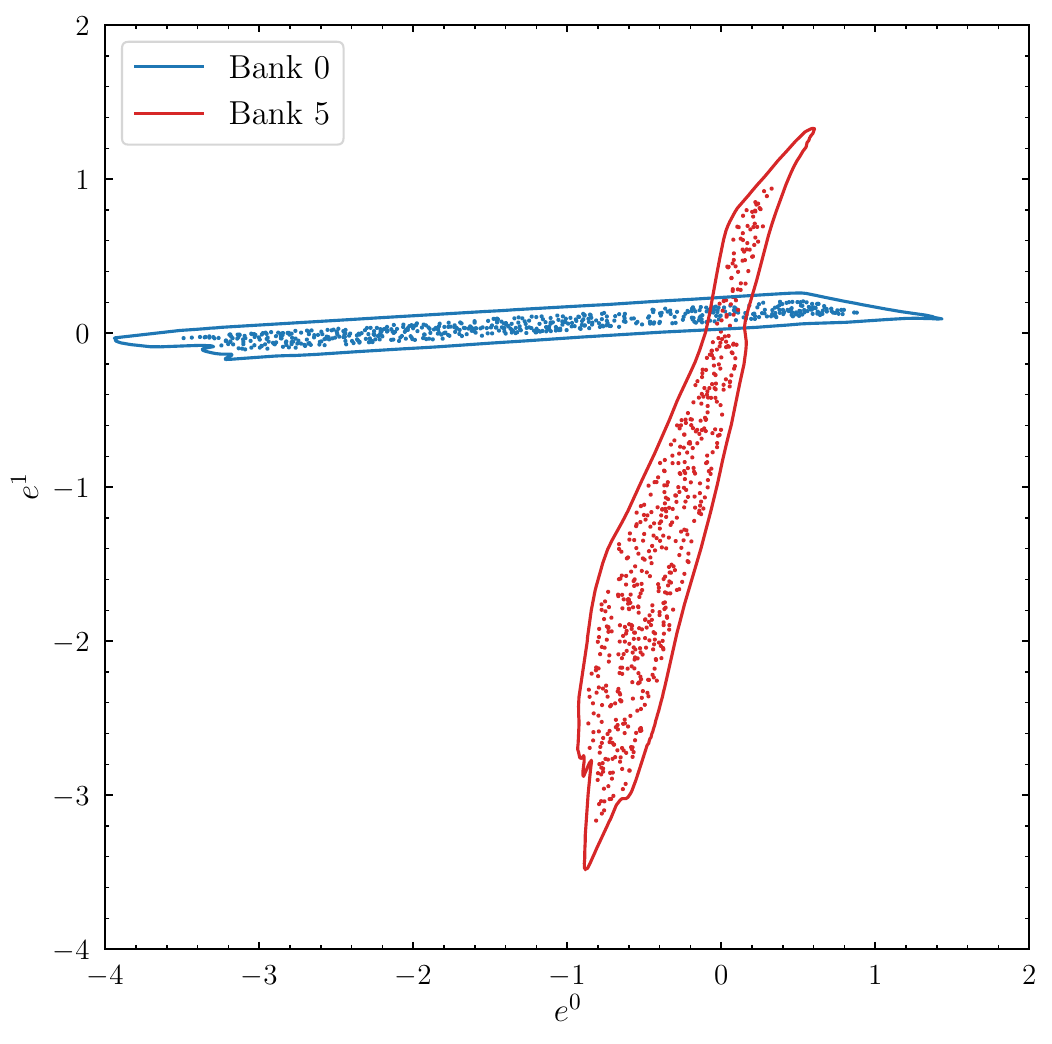} 
\includegraphics[width = 0.495\textwidth]{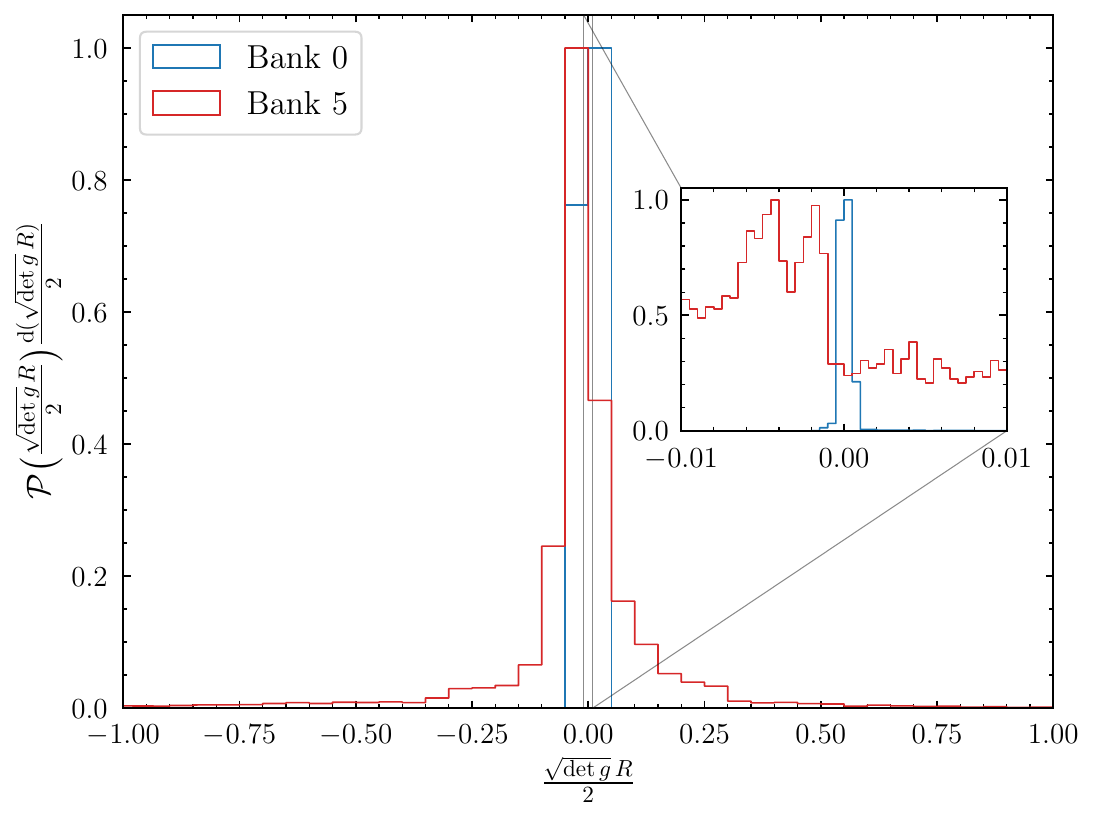} 
\caption{The left panel shows the $99.7\%$ CL contours of the distribution of the latent-space points (the encoding of the SVD components $c^A$) for Bank $0$ and Bank $5$, together with $5\%$ of the $12288$ points used to compute the curvature of the banks (notice that in this plot we use the autoencoders and normalizing flows refined to \texttt{torch.float64} precision). The right panel shows the distribution of $\smash{\sqrt{g}\,R/2}$ for the two banks (in the case of Bank $5$, we drop $698$ points that have estimated $\smash{|\sqrt{g}\,R/2|}$ significantly larger than $1$ to allow the two histograms to be shown in the same plot).}
\label{fig:autoencoder_curvature}
\end{figure}

\noindent In Fig.~\ref{fig:GWs_autoencoder_distance_preservation_b_0_5} we see that the bank $\text{BBH-5}, {\cal M}\in[14.7,23.9]\,{M_\odot}$ (``Bank $5$'' hereafter) has a bigger scatter in the distance preservation than the other three banks discussed in the main text. We also checked that this bank is the only one to show this behavior across all banks and sub-banks. We can then compute the curvature of the bank in a similar way as in Appendix~\ref{app:sphere}. Since in this case we do not have a target curvature, instead of estimating the Ricci scalar we look at 
\be
\sqrt{g}\,R\sim\partial\partial\ln g+(\partial\ln g)^2\,\,.
\ee
In this way, we are insensitive to the overall size of the metric in $2$ dimensions and we have a fair estimate of the curvature radius (recall that the Riemann tensor is essentially of the form $\smash{\partial\partial\ln g+(\partial\ln g)^2}$). We proceed in the same way as for the sphere: our estimator for the curvature is
\be
\label{eq:curvature_estimator_with_determinant}
\frac{\sqrt{g(e)}\,R(e)}{2}\approx\frac{\sqrt{g(e)}\,P^1_{\hphantom{1}0}(\partial D_\epsilon(e))}{\int_{D_\epsilon(e)}{\rm d}^2e'\,g_{00}(e')}\,\,,
\ee
where again the holonomy is computed around a circle $\partial D_\epsilon(e)$ of radius $\epsilon$ centered in $e$. 

We choose $12288$ physical points for the two banks, making sure that they are at least a distance of $0.05$\footnote{Notice that this is intended as Euclidean distance in $2$ dimensions, consistently with the formulation of the non-Abelian Stokes theorem.} from the $99.7\%$ CL contour in order to be sure that all the circles around which we compute the holonomy are also inside this contour. We then carry out the same procedure as in Appendix~\ref{app:sphere} to compute the numerator and denominator of Eq.~\eqref{eq:curvature_estimator_with_determinant}. The results are shown in Fig.~\ref{fig:autoencoder_curvature}: we see that while both banks have a distribution of the curvature consistent with zero, the higher-mass bank has a spread in curvature which is around two orders of magnitude larger than the lower mass bank. This was expected from what we saw in Fig.~\ref{fig:GWs_autoencoder_distance_preservation_b_0_5}. Also, it is important to emphasize that the curvature is an invariant of the manifold: hence this result is completely independent of the choices of $\alpha,\beta,\gamma$ we use when training the autoencoders (as long as both of them describe the respective manifolds with a comparable reconstruction error).

\begin{figure}[t]
\centering
\includegraphics[width = 0.7\textwidth]{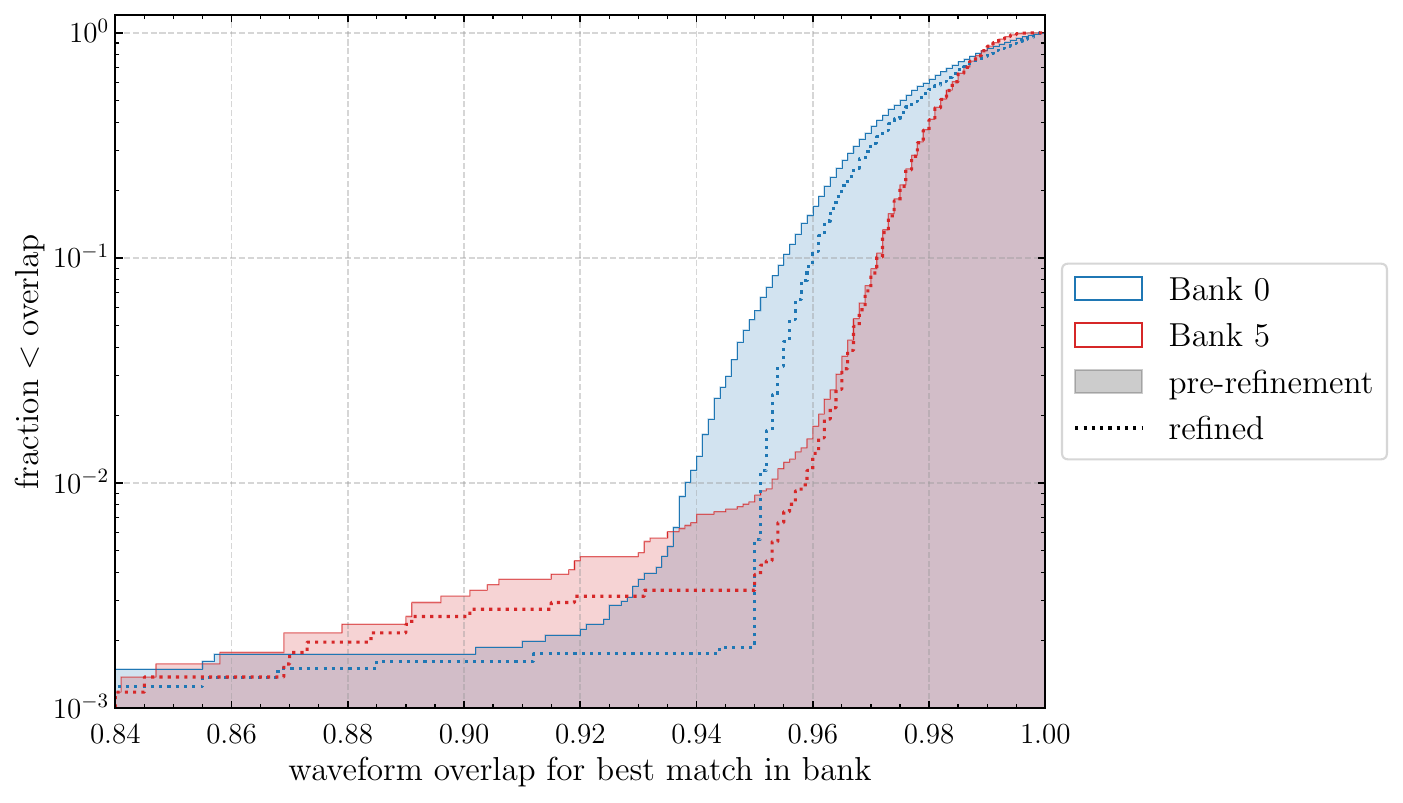} 
\caption{Effectualness of the template banks for Bank $0$ and Bank $5$. We see that the higher-mass (and higher-curvature) one benefits from a grid refinement slightly less than the lower-mass one.} 
\label{fig:GWs_effectualness_curvature}
\end{figure}

This difference in the spread of curvature notwithstanding, Fig.~\ref{fig:GWs_effectualness_curvature} shows that the effectualness tests for both banks yield comparable results, even if for Bank $5$ we see that even after grid refinement a $0.3\%$ fraction of test waveforms has an overlap with their best match in the bank less than $0.95$, at variance with Bank $0$ for which this fraction is closer to $0.1\%$. Finally, we notice that the two normalization factors $\cal N$ are respectively equal to $55.6$ and $3.4$ for the two banks: hence in the first case we have a grid size of order $\Delta e = 0.55/{\cal N}\approx 10^{-2}$, and in the second case of order $\Delta e = 0.55/{\cal N}\approx 10^{-1}$. If we estimate by eye the spread of the inverse of curvature radius of the first bank to be $2\times10^{-3}$, and $0.5$ for the second bank, we see that the grid size is many orders of magnitude smaller than the curvature radius for the lower-mass bank compared to the higher-mass bank. We do not comment more on the relation between grid size and bank curvature in this paper, leaving a more detailed investigation to future work.

\subsection{\texorpdfstring{Autoencoder for \texttt{COBRA}}{Autoencoder for COBRA}}
\label{app:preliminary}

\begin{figure}[ht]
    \centering
    \settoheight{\FigHeight}{\includegraphics[width=0.99\textwidth]{plots_sphere/points_for_curvature_calculation.png}}
    \begin{subfigure}{\textwidth}
        \centering
        \includegraphics[width = 0.495\textwidth]{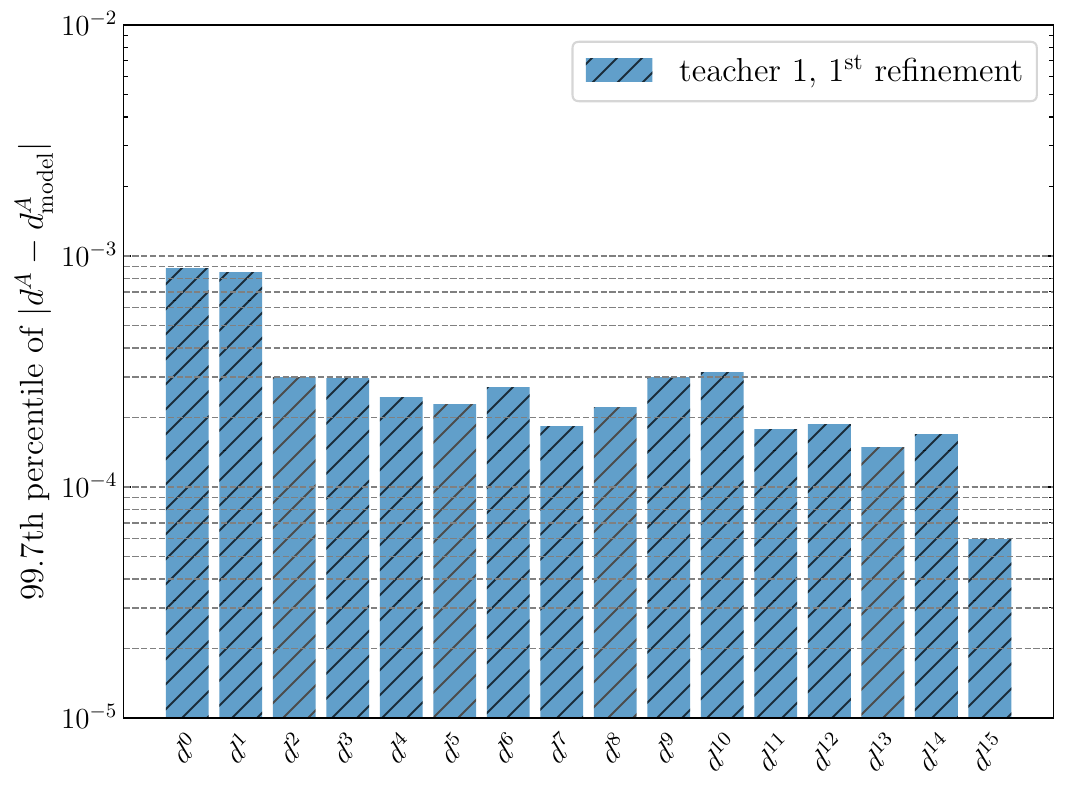} 
        \includegraphics[width = 0.495\textwidth]{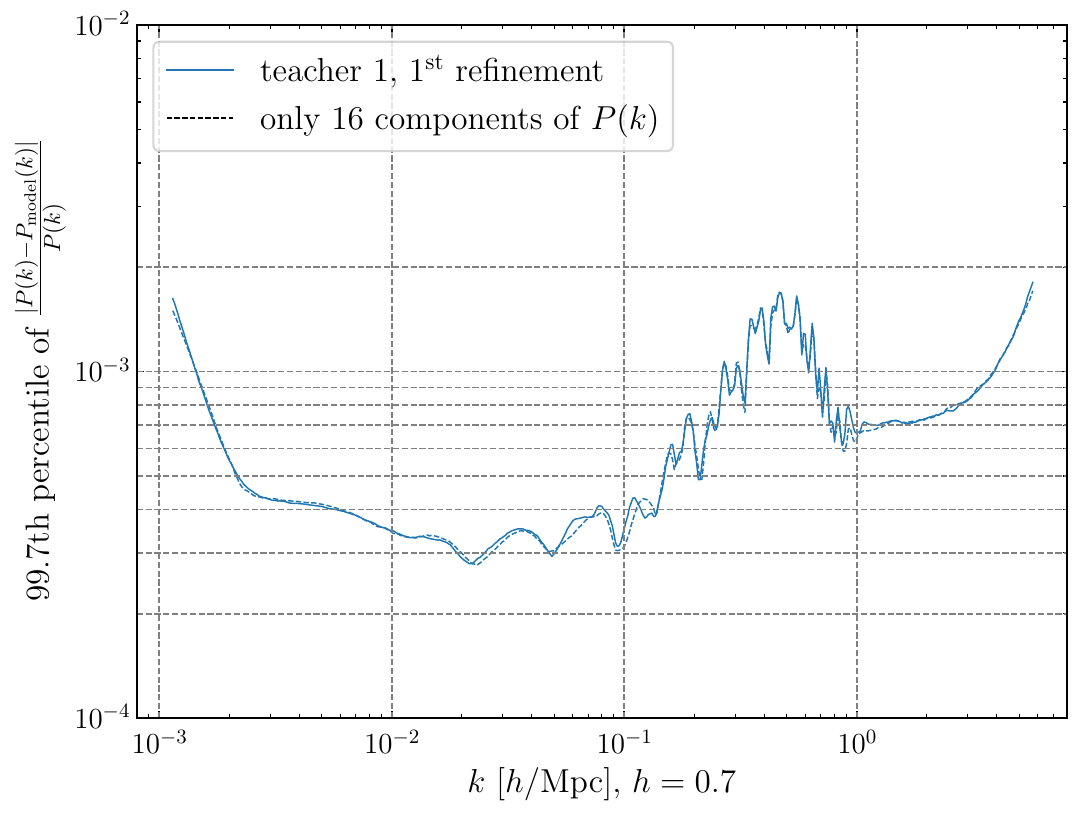} 
        \caption{\small Same as Fig.~\ref{fig:cobra}, but for the autoencoder architecture presented in Eq.~\eqref{eq:cobra_autoencoder_architecture}. We show only the result of a preliminary pass, that can be considered just as the training of a first ``teacher'' for a future ensemble + transfer learning procedure, in case that more precision is needed.}
        \label{fig:cobra_autoencoder-upper}
    \end{subfigure}
    \hfill
    \begin{subfigure}{\textwidth}
        \centering
        \resizebox{!}{\FigHeight}{\includegraphics[width=\textwidth]{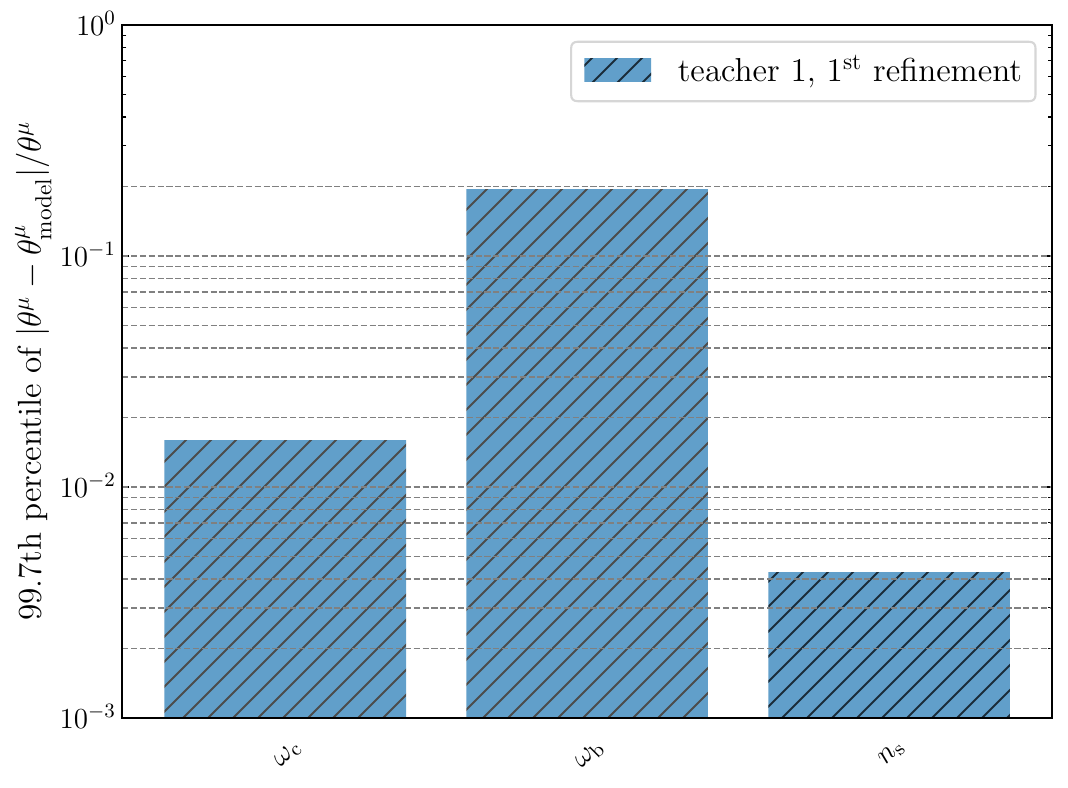}}
        \caption{\small $99.7$th percentile of the relative error in the validation cosmological parameters reconstructed via the $\smash{d^A\to e^\mu\to\theta^\mu}$ transformation described in Eq.~\eqref{eq:cobra_autoencoder_architecture}. We see that it is $\omega_{\rm b}$ in particular that is difficult to reconstruct.}
        \label{fig:cobra_autoencoder-lower}
    \end{subfigure}
\invisiblecaption{fig:cobra_autoencoder}
\vspace*{-\baselineskip}
\end{figure}

\noindent In Section~\ref{sec:conclusions} we discussed the idea of constructing an autoencoder for the SVD decomposition of the linear power spectrum of Ref.~\cite{Bakx:2024zgu}. In Fig.~\ref{fig:cobra} we have shown what we could call the decoder in this setup. We just trained a MLP to reproduce the $c^A$ in terms of the cosmological parameters $\theta=(\omega_{\rm c},\omega_{\rm b},n_{\rm s})$, reaching a precision slightly better than the RBF interpolation of \cite{Bakx:2024zgu}. If we want to actually obtain a full autoencoder, we also need to construct and train the encoder, with the constraint that now we want that the latent space is the space of $\theta^\mu$. To this purpose, we constructed an encoder with a $16$-dimensional input layer, $8$ hidden layers each containing $128$ neurons and a GELU activation, and a $3$-dimensional output layer to go from the $d^A$, defined as $c^A/{\cal N}$, to the latent space of some variables $e^\mu$. Then, instead of forcing via a simple ${\rm MSE}$ loss that $e^\mu\approx\theta^\mu$, in order to add flexibility we constructed an invertible transformation $f^\mu$ using \texttt{zuko} with the goal that $\theta^\mu=f^\mu(e)$. More precisely, $f^\mu$ is a composition of a sequence of $24$ \href{https://zuko.readthedocs.io/stable/api/zuko.flows.autoregressive.html#zuko.flows.autoregressive.MaskedAutoregressiveTransform}{Masked Autoregressive Transforms} ($4$ repetitions of $3!=6$ permutations) that utilize Monotonic Affine Transforms as their univariate components. The parameters for the Monotonic Affine Transform are predicted by an internal neural network architecture consisting of $4$ hidden layers each with $64$ neurons and a GELU activation function. These transforms are followed by the composition with an untrainable sigmoid transform and finally with an affine transformation acting separately on each $e^\mu$ in order to bring them inside the box covered by the uniform and structured grid used in Ref.~\cite{Bakx:2024zgu}. In other words, our architecture is 
\begin{equation}
\label{eq:cobra_autoencoder_architecture}
\smash{c^A\overset{\raisebox{-1.5em}{${\scriptstyle \cal E}$}}{\longrightarrow} e^\mu \overset{\raisebox{-1.5em}{${\scriptstyle f}$}}{\longrightarrow} \theta^\mu \overset{\raisebox{-1.5em}{${\scriptstyle \cal D}$}}{\longrightarrow} c^A}\,\,.
\end{equation}
As a first pass, we train both the ``flow'' $f^\mu$ and the autoencoder -- the first teacher for a possible transfer learning to improve precision at a later stage -- with the loss being a sum of three terms (each with the same overall weight): an MSE loss for $\smash{d^A - \tilde{d}^A_{\rm train}}$ (where $\smash{\tilde{d}^A_{\rm train}}$ is again defined by $\smash{{\cal D}^A({\cal E}({d}_{\rm train}))}$, like in the training of the autoencoders in the main text), one for $\theta^\mu - f^\mu(e)$, and one for $d^A - {\cal D}^A(p)$. The decoder weights are initialized to those of the one shown in orange in Fig.~\ref{fig:cobra}. The results of this preliminary first pass are shown in Fig.~\ref{fig:cobra_autoencoder}: while it is clear that more refinement is needed in order both reproduce the precision of Fig.~\ref{fig:cobra} in the $\theta^\mu\to d^A \to P(k)$ mapping, and the precision in our recovery of cosmological parameters $P(k)\to d^A\to e^\mu\to\theta^\mu$ is not optimal -- especially in the case of $\omega_{\rm b}$ -- our architecture is clearly able to find an ``inverse'' of \texttt{CAMB} in this particular setup where we vary $(\omega_{\rm c},\omega_{\rm b},n_{\rm s})$. Notice also that it is possible our architecture is not the bottleneck in getting from spectra to cosmological parameters, since from the right panel of Fig.~\ref{fig:cobra} we see that the improvement in precision after we refine our decoder does not translate in a similar improvement on the spectra, likely suggesting that we are dominated by the error we are making when we neglect SVD components beyond $A=15$, that seem to be needed to capture the BAO wiggles between $0.1\,h/{\rm Mpc}$ and $1\,h/{\rm Mpc}$ to the precision of $\sim 10^{-4}$.

We conclude by pointing out that we were motivated to use the invertible function $f^\mu$ since in principle, had we trained the autoencoder only with a reconstruction loss forcing $e^\mu\approx\theta^\mu$, it would have been difficult for it to respect the distribution of cosmological parameters in our training set. It should be possible to enforce the distribution of $e^\mu$ to be such that the distribution of $\theta^\mu$ is uniform -- i.e.~to ask $f^\mu(e)$ to match also the density of $\theta^\mu$ instead of their position only -- by turning on the loss 
\begin{equation}
\label{eq:NF_loss}
\bigg\langle{-\ln {\cal P}_\theta\big(f(e)\big)}\cdot\bigg|\frac{\partial f(e)}{\partial e}\bigg|\bigg\rangle_e\,\,,
\end{equation}
where $\smash{{\cal P}_\theta}$ is the uniform distribution inside the box used in Ref.~\cite{Bakx:2024zgu}. This could be a next step once we move on from a training set consisting of points on a uniform and structured grid (for which the distribution is more akin to a sum of Dirac delta functions). Moreover, having the function $f$ could be useful if later we want to impose the latent space to have some specific property, e.g.~distance preservation, or as a starting point for the training of a variational autoencoder to learn the posterior on cosmological parameters.

\bibliography{reference}

\end{document}